\documentclass[12pt]{article}
\usepackage{graphicx} 
\usepackage{amsmath}

\usepackage{adjustbox}
\usepackage{makecell}
\usepackage{rotating}
\usepackage{amssymb}
\usepackage{arydshln}
\usepackage{arxiv}

\usepackage[utf8]{inputenc} 
\usepackage[T1]{fontenc}    
\usepackage{hyperref}       
\usepackage{url}            
\usepackage{booktabs}       
\usepackage{amsfonts}       
\usepackage{nicefrac}       
\usepackage{microtype}      
\usepackage{cleveref}       
\usepackage{natbib}
\usepackage{doi}

\title{A framework for parametric time-varying treatment effects in multiple intervention stepped wedge design clinical trials in the presence of non-uniform cluster-period correlation structures}

\author{ \href{https://orcid.org/0000-0002-8663-5079}{\includegraphics[scale=0.06]{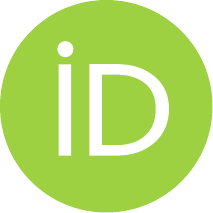}\hspace{1mm}Samantha M.~Levy, PhD}\\
	Asymmetric Operations Sector\\
	The Johns Hopkins University Applied Physics Laboratory\\
	Laurel, Maryland \\
	\texttt{samantha.levy@jhuapl.edu} \\
	\And
	\href{https://orcid.org/0000-0003-2505-0090}{\includegraphics[scale=0.06]{orcid.pdf}\hspace{1mm} Jose-Miguel~Yamal, PhD} \\
	Department of Biostatistics and Data Science\\
 UTHealth Houston School of Public Health, The University of Texas Health Science Center at Houston\\
	Houston, Texas \\
	\texttt{Jose-Miguel.Yamal@uth.tmc.edu} }

\renewcommand{\shorttitle}{Parametric Time Varying Treatment Effects in M-SWDs}

\hypersetup{
pdftitle={Parametric Time Varying Treatment Effects in M-SWDs},
pdfsubject={stat.ME},
pdfauthor={Samantha M.~Levy, Jose-Miguel~Yamal},
pdfkeywords={stepped wedge design, multiple intervention trials, time-varying treatment effects, power and sample size, model misspecification},
}

\begin{document}
\maketitle

\begin{abstract}
	Stepped wedge design (SWD) trials typically assume that treatment effects are immediate following implementation, but this assumption is often unrealistic in pragmatic settings where effects evolve over exposure time. This impact is further complicated in multiple-intervention stepped wedge designs (M-SWDs), due to their complex crossover patterns. Existing work has addressed time-varying treatment effects for main effects at the analysis stage using nonparametric approaches; however, the implications for design-stage power and estimand interpretation in M-SWDs remain largely unaddressed. We develop a unified framework that incorporates parametric exposure-response functions into a modified fixed-effects design matrix, $\mathbf{Z}^*$, and derive corresponding generalized least squares (GLS) expressions for treatment effect estimands, variance, power, and bias under model misspecification. Analytic and simulation results demonstrate that time-varying effects can require multiple-fold increase in the number of clusters required to achieve the nominal target power. Misspecification can also induce large bias in treatment effect estimates, with particularly pronounced and non-uniform impacts on interaction effects. These findings demonstrate that treatment effect estimands in M-SWDs depend critically on assumptions about exposure-time dynamics and that failure to account for time-varying effects can lead to substantial miscalibration of power and biased inference, affecting main and interaction effects differently. 
\end{abstract}

\keywords{stepped wedge design, multiple intervention trials, time-varying treatment effects, power and sample size, model misspecification}

\section{Introduction}
\begin{figure}
    \centering
    \includegraphics[width=0.65\linewidth]{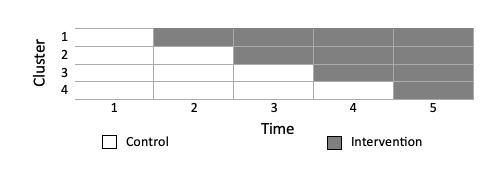}
    \caption[Simple Stepped Wedge Design]{Simple SWD with a single intervention, 4 clusters, and 5 periods.}
    \label{fig:simpleswd}
\end{figure}

In stepped wedge design (SWD) clinical trials, typically each cluster begins in the control phase and then crosses to the intervention phase such that, by the final period, all clusters receive the intervention as shown in Figure \ref{fig:simpleswd}. SWDs are especially useful when the intervention is expected to be universally implemented by the end of the study or when researchers must evaluate the impact of an intervention while simultaneously implementing a large-scale roll-out \citep{piszczek_stepped-wedge_2015, brown_stepped_2006, federico_ethical_2022, cook_statistical_2016}. 

The staggered roll-out in SWDs can introduce multiple sources of bias, particularly those arising from time trends and longitudinal correlation. Accounting for time effects and realistic correlation structures is essential to avoid biased treatment effect estimates and standard errors \citep{thompson_bias_2017, kasza_inference_2019}. The traditional Hussey-Hughes linear mixed model (LMM) assumes a single, constant intracluster correlation (ICC) and an instantaneous treatment effect \citep{hussey_design_2007}. Subsequent work has shown that these assumptions are often unrealistic and can lead to substantial bias, especially in pragmatic settings where complex temporal dependencies or delayed intervention uptake are common \citep{kasza_impact_2019, hussey_design_2007, hughes_current_2015}. Specifically, since treatment effects are typically calculated through generalized least squares (GLS) estimation in SWDs, any misspecification of the within-cluster correlation structure can alter the covariance structure, while misspecification of the temporal treatment effects can alter the implicit weighting of the observations over time, leading to a compound effect on bias. 

A major concern in modeling SWDs is the presence of time-varying treatment effects, specifically the scenario in which the full treatment effects are not instantaneous at transition to the intervention phase \citep{hughes_current_2015}. To address this bias, it is important to consider the impact of exposure time - the time since the cluster transitioned from the control to the intervention \citep{nickless_2018, kenny_2022, hughes_current_2015}. These time-varying treatment effects can be modeled using "effect curves" such that the magnitude of the treatment effect is expressed as a function of exposure time. Time-varying treatment effect curves must be considered during both the design and analysis phase, with design phase favoring parametric approaches and analysis phase favoring non-parametric approaches 
\citep{hughes_current_2015, kenny_2022}. Failing to model the effect curve correctly can bias estimates so significantly that an intervention with a true positive effect may appear neutral or even harmful \citep{kenny_2022}. 

\begin{figure}
    \centering
    \includegraphics[width=0.5\linewidth]{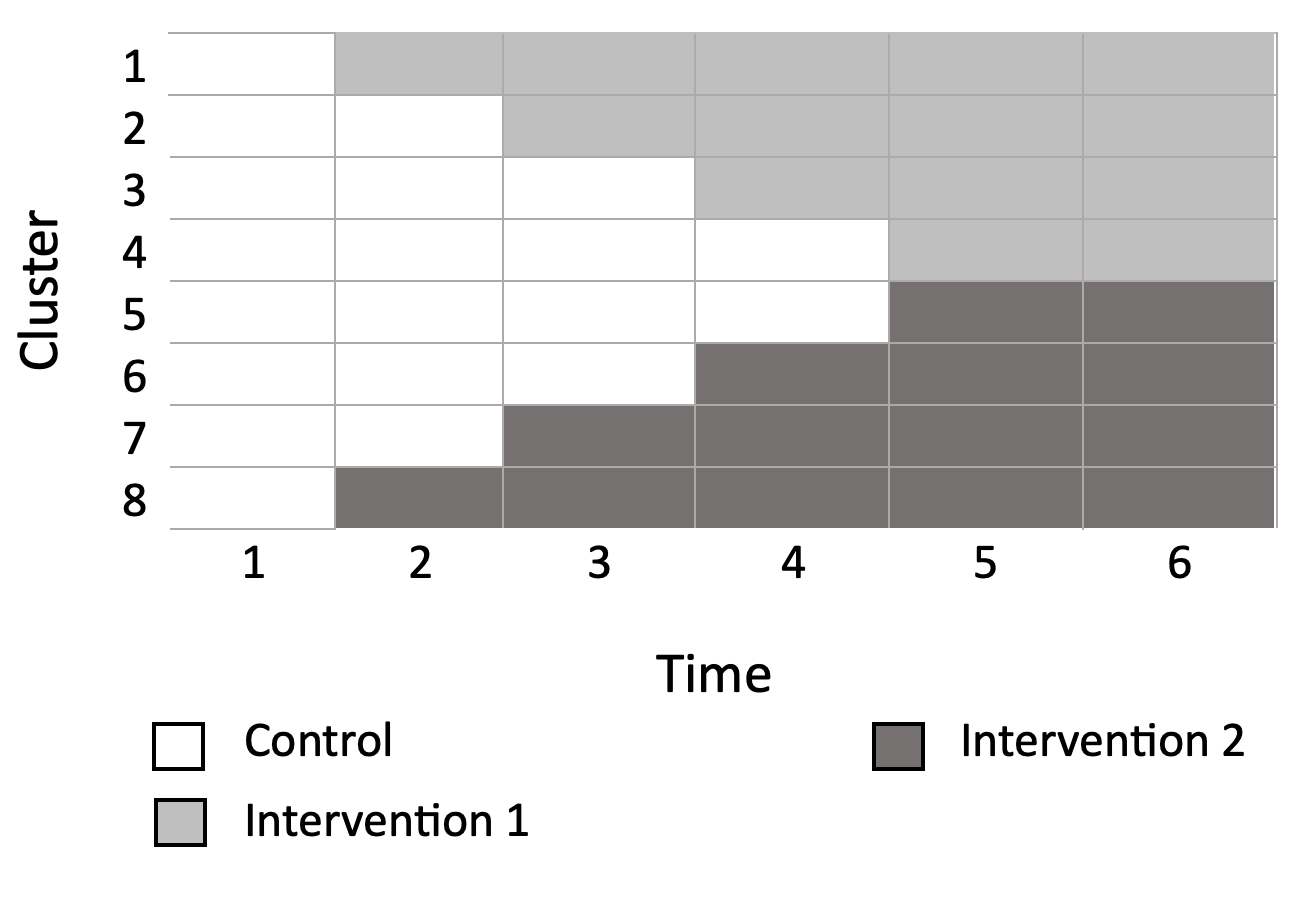}
 \caption[Concurrent Multiple Intervention Stepped Wedge Design]{Example concurrent M-SWD study for 8 clusters across 6 periods.  White cluster period cells are in the control period, light gray are receiving only intervention 1, and mid-gray are receiving only intervention 2.}
    \label{tab:cont1}
\end{figure}

\begin{figure}
    \centering
    \includegraphics[width=0.5\linewidth]{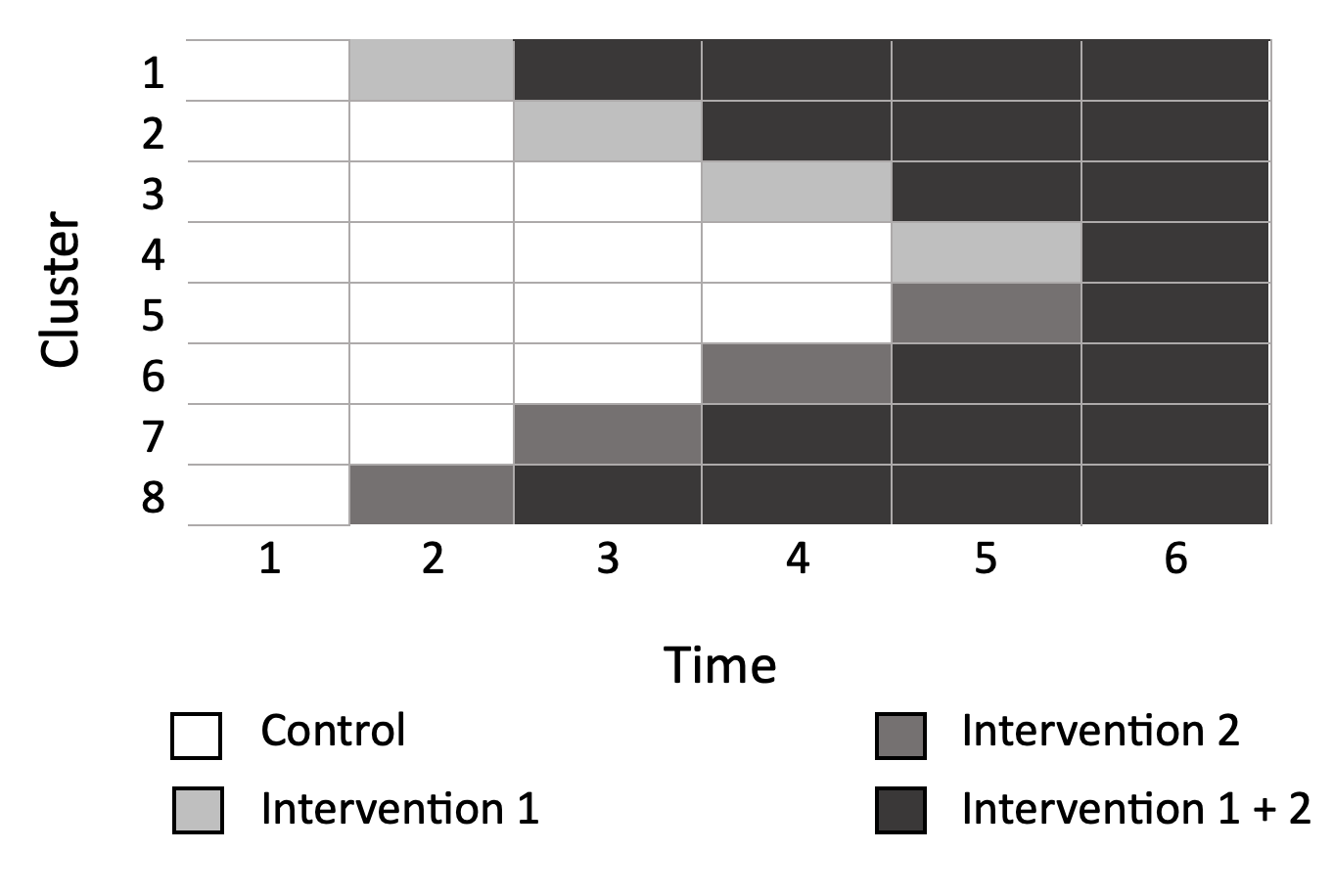}
    \caption[Factorial Multiple Intervention Stepped Wedge Design]{Example factorial M-SWD study for 8 clusters across 6 periods. White cluster period cells are in the control period, light gray are receiving only intervention 1, mid-gray are receiving only intervention 2, and dark gray receiving both interventions 1 and 2.}
    \label{tab:fact1}
\end{figure}

A more recent development in stepped wedge design trials is the multiple-intervention SWD trial (M-SWD), in which two or more interventions are compared to a common control within one stepped wedge framework \citep{lyons_proposed_2017}. Sample M-SWDs are shown in Figures \ref{tab:cont1} and \ref{tab:fact1}. Although M-SWDs are increasingly implemented in practice, statistical methods for their design and analysis remain limited \citep{sundin_power_2022, pol_effectiveness_2017, pol_effectiveness_2019, finney_rutten_evaluating_2018, zhu_enhancing_2023}. \cite{sundin_power_2022} extended the standard LMM framework to accommodate multiple interventions with an interaction for continuous outcomes, but assumed compound-symmetry correlation and immediate treatment effects \citep{sundin_power_2022}. This work has been recently expanded to allow for flexible cluster-period structures \citep{paper1arxiv}. More recent studies have generalized models to estimate non-parametric time-varying effects post hoc \citep{kenny_2022, chen_timevar_2025}, but these approaches focus on data-driven analysis-stage estimation of effects rather than design-stage power.

To address these limitations, this work develops a generalized framework for M-SWDs that allows for parametric time-varying treatment effects and flexible cluster-period correlation structures designed for utilization in design-stage power calculations. Building on \cite{sundin_power_2022} and on the correlation framework established in \cite{paper1arxiv}, this framework models non-instantaneous parametric treatment effects, constructs structured covariance and design matrices, and provides analytic expressions towards power, variance, and bias under these formulations. The remaining sections introduce notation and GLS estimation (\S\ref{sec:aim2note}-\S\ref{sec:GLS}); present parametric effect curves and time-varying interaction surfaces (\S\ref{sect:param}-\S\ref{sec:timesurf}); develop power, bias, and variance formulas (\S\ref{sec:power}-\S\ref{sec:bias}); and extend to random-effects models (\S\ref{sec:randomeffect}). We also design an analytic and simulation study to validate and assess these results (\S\ref{sec:simaim2}) and present these results showing the impact of time-varying treatment response functions (\S\ref{sec:simaim2res}). We conclude with a discussion of findings, limitations, and directions for future research (\S\ref{sec:simaim2discuss}).

\section{Methodology}\label{sec:methaim1}

\section{Methodology}

\subsection{Notation and Base Model} \label{sec:aim2note}

Assuming additive and immediate treatment effects, the multiple intervention stepped wedge design model for $Q$ interventions is:

\begin{equation} \label{eq:fullestsimp}
    Y_{ijk} = \mu + \beta_j +  \alpha_i + \nu_{ij} + \psi_{ik} + \sum_{q=1}^Q \theta_q X_{qij} + e_{ijk}
\end{equation}
where $Y_{ijk}$ is the individual level response for $i = 1, \ldots, I$ clusters, $j=1, \ldots, T$ time points, $k=1, \ldots, N$ individuals. $\theta_q$ is the fixed treatment effect for treatment $q$ and $X_{qij}$ is an indicator variable defined to equal 1 when cluster $i$ receives treatment $q$ at time $j$ and 0 when cluster $i$ does not receive treatment $q$ at time $j$. Under the simplest case, $Q=1$, this model reduces to the standard one treatment model. $\alpha_i$ is the random effect for cluster $i$ such that $\alpha_i \sim N(0, \sigma^2_\alpha)$, $\nu_{ij} \sim N(0, \sigma^2_\nu)$ is the random effect for cluster $i$ at time $j$, $\psi_{ik}$ is the random effect for individual $k$ in cluster $i$ such that $\psi_{ik} \sim N(0,\sigma^2_\psi)$, $\beta_j$ is a fixed effect for time period $j$, such that for identifiability $\beta_T = 0$, and $e_{ijk} \stackrel{iid}{\sim} N(0, \sigma^2_e)$ is the marginal variance. The total variance of an individual response, $Y_{ijk}$, is therefore:
\begin{equation}\label{eq:aim2totalvar}
\text{Var}(Y_{ijk})= \sigma^2_y = \text{Var}(\alpha_i) + \text{Var}(\nu_{ij}) + \text{Var}(\psi_{ik}) + \text{Var}(e_{ijk}) = \sigma^2_\alpha + \sigma^2_\nu + \sigma^2_\psi + \sigma^2_e.   
\end{equation}

A common multiple-intervention setting is the case of two interventions with an optional interaction:
\begin{equation}\label{eq:fulltimevar}
    Y_{ijk} = \mu + \beta_j +  \alpha_i + \nu_{ij} + \psi_{ik} + \theta_1 X_{1ij} + \theta_2 X_{2ij} + \theta_3 X_{1ij}X_{2ij} + e_{ijk} .
\end{equation}

Models of the form in Equations \eqref{eq:fullestsimp} and \eqref{eq:fulltimevar} assume that treatment effects are immediate and constant following crossover. This assumption may not hold in practice when intervention effects depend on exposure duration. Previous work has examined time-varying treatment effects in stepped wedge designs for data-driven post-hoc estimation under both single-intervention and multiple-intervention settings \citep{hughes_current_2015, kenny_2022, chen_timevar_2025}. In particular, Kenny et al. \citep{kenny_2022} and Chen et al. \citep{chen_timevar_2025} represent time-varying effects nonparametrically through exposure-time-specific parameters indexed by cumulative exposure. We build on this idea, with a focus on design-stage parametric formulations for power and sample size calculations.

\subsection{Cluster-Period Mean}\label{sec:aim2CP} 
We define the cluster-period mean model:

\begin{equation}
    \bar{Y}_{ij.} = \mu + \beta_j + \alpha_i + \nu_{ij} + \psi_{i.} + \sum_{q=1}^Q \theta_{q}X_{qij} + e_{ij.}
\end{equation}
where $\psi_{i.} = \frac{1}{N}\sum_{k=1}^N\psi_{ik}\sim N(0,\sigma^2_\zeta)=N(0,\sigma^2_\psi/N)$ is the cluster-level average of the individual random effects. Similarly, $e_{ij.}=\frac{1}{N}\sum_{k=1}^Ne_{ijk} \sim N(0,\sigma^2_c)=N(0,\sigma^2_e/N)$. The total cluster-period variance is:
\begin{equation}\label{eq:aim2eta}
    \eta = \sigma^2_\alpha + \sigma^2_\nu + \sigma^2_\zeta + \sigma^2_c.
\end{equation}

\subsection{Generalized Least Squares (GLS) Framework}\label{sec:GLS}
\subsubsection{Formulation of Standard Fixed Effects Design Matrix, $Z$}\label{sect:Z_s}
We first define the fixed-effects parameter vector:
\begin{equation*}
\boldsymbol{\Theta}=   \begin{bmatrix}
    \mu & \beta_1 & \ldots & \beta_{T-1} & \theta_1 & \ldots & \theta_Q 
\end{bmatrix}'.
\end{equation*}
The standard fixed effect design matrix is  $\mathbf{Z} \in \mathbb{R}^{IT \times (T + Q)}$ as:
\begin{equation*}
\mathbf{Z} = 
\begin{bmatrix}
    \mathbf{Z}_1 \\
    \mathbf{Z}_2 \\
    \vdots \\
    \mathbf{Z}_I
\end{bmatrix}
\end{equation*}
with each $\mathbf{Z}_i \in \mathbb{R} ^{T \times (T+Q)}$ of the form:
\renewcommand{\arraystretch}{0.4}
\begin{equation} \label{eq:zfull}
\mathbf{Z}_i = 
\begin{bmatrix}
    & \mathbf{I}_{T-1} & & &  \\   \mathbf{1}_T && \mathbf{X}_{1i} & \ldots & \mathbf{X}_{Qi} \\
   & \mathbf{0}'_{T-1} & & &
\end{bmatrix}.
\end{equation}
Each $\mathbf{X}_{qi} = (X_{qi1}, X_{qi2}, \ldots, X_{qiT} )'$ denotes the treatment indicators for intervention $q$ in cluster $i$. The $\mathbf{I}_{T-1}$ matrix contains indicators for times $1, \ldots T-1$, corresponding to fixed time effects, and $\mathbf{0}'_{T-1}$ corresponds to time $T$ under identifiability constraint $\beta_T=0$.

Under the common case of two interventions with an interaction, this simplifies to:

\begin{equation}\label{eq:Zstand}
\mathbf{Z}_i = 
\begin{bmatrix}
    & \mathbf{I}_{T-1} & & &  \\   \mathbf{1}_T && \mathbf{X}_{1i} & \mathbf{X}_{2i} & (\mathbf{X}_1 \mathbf{X}_2)_i \\
   & \mathbf{0'_{T-1}} & & &
\end{bmatrix}.
\end{equation}
We can partition the design matrix as:
\begin{equation*}
    \mathbf{Z} = \begin{bmatrix}
        \mathbf{Z}^{(\text{time})} &  \mathbf{Z}^{(\text{1})} &  \mathbf{Z}^{(\text{2})} &  \mathbf{Z}^{(\text{1,2})}
    \end{bmatrix}
\end{equation*}
corresponding to calendar-time effects, the two main intervention effects, and the interaction effect, respectively. 

\subsubsection{Structured Cluster-Period Random Effects in Time-Varying Treatment Models}\label{sect:nu_aim2}
Structured cluster-period random effects are modeled as:
\begin{equation*}
    \boldsymbol{\nu}_i=(\nu_{i1}, \ldots , \nu_{iT})^T \sim N(0, \sigma_\nu^2R)
\end{equation*}
where $R=[r_{jm}]_{T\times T}$ is a correlation matrix with structure $r_{jj}=1$ and $|r_{jm}|\leq 1$ for all $j,m$, following from \cite{paper1arxiv}. Examples include compound symmetry, autoregressive order 1 (AR(1)), proportional decay, and Toeplitz structures. This specification determines the cluster-period covariance structure used in the GLS framework below.

\subsubsection{GLS Framework for Estimands}\label{sect:GLS}
We work at the cluster-period mean level, defining:
\begin{equation*}
    \mathbf{\bar{Y}_{i}} = (\bar{Y}_{i1.}, \ldots, \bar{Y}_{iT.})^T.
\end{equation*}
The variance-covariance structure of $Y_i$ follows from the random effect structure in \S\ref{sect:nu_aim2} and $\bar{Y}_{ij}$ follows from Section \S\ref{sec:aim2CP}. The total cluster-period mean variance remains as in equation \eqref{eq:aim2eta} and the covariance for cluster $i$ between period $j$ and $m$ is:
\begin{equation*}
    \text{Cov}(\bar{Y}_{ij}, \bar{Y}_{im}) = \sigma^2_\alpha + \sigma^2_\nu r_{jm} + \sigma^2_\zeta, \hspace{16pt} j\neq m
\end{equation*}
where $R=[r_{jm}]$ is the correlation matrix for the cluster-period random effects.
This yields a cluster-specific covariance matrix:
\begin{equation*}
    \mathbf{V}_i=\sigma^2_\nu R + \sigma^2_cI_T + (\sigma^2_\alpha + \sigma^2_\zeta)J_T 
\end{equation*}
Assuming independence across clusters, the full outcome variance-covariance matrix is block diagonal:
\begin{equation*}
    \mathbf{V} = \text{diag}(\mathbf{V}_1, \ldots, \mathbf{V}_I).
\end{equation*}
For a design matrix $\mathbf{Z}$ and fixed effect vector $\mathbf{\Theta}$, the GLS estimator is:
\begin{equation}
    \hat{\boldsymbol{\Theta}}= (\mathbf{Z}^T\mathbf{V}^{-1}\mathbf{Z})^{-1}\mathbf{Z}^T\mathbf{V}^{-1}Y,
\end{equation}
with expected value: 
\begin{equation}
    E[\boldsymbol{\hat\Theta}] = (\mathbf{Z}^T\mathbf{V}^{-1}\mathbf{Z})^{-1}\mathbf{Z}^T\mathbf{V}^{-1}E[Y] = (\mathbf{Z}^T\mathbf{V}^{-1}\mathbf{Z})^{-1}\mathbf{Z}^T\mathbf{V}^{-1}\mathbf{Z}\boldsymbol{\Theta}=\boldsymbol{\Theta},
\end{equation}
and covariance:
\begin{equation}
     \text{Cov}(\hat{\boldsymbol{\Theta}})= (\mathbf{Z}^T\mathbf{V}^{-1}\mathbf{Z})^{-1}.
\end{equation}

\subsection{Parametric Exposure-Response Functions}\label{sect:param}
Previous work has represented exposure-time effects nonparametrically, using a separate parameter, $\delta_{q,e}$ for each exposure period, $e$. While this is useful for data-driven post-hoc estimation, it is often necessary to estimate the impact of a time-varying treatment effect during the design stage. We therefore define $s_{qij} \in \{0,1,\ldots,T-1\}$ as the number of periods since cluster $i$ transitioned to intervention $q$ at the start of interval $[j,j+1]$, and define all time-varying regressors as period averages over that interval. Throughout, we assume a uniform observation density within each cluster-period.

We therefore introduce parametric response curves, also known as effect curves:
\begin{equation}
\theta_q(s_q)=\theta_q^{\text{max}}f_q(s_q;\phi_q),
\end{equation}
where each $f_q(s_q;\phi_q)$ is a function of 1 - 3 interpretable parameters.

A full model for $Q$ interventions under the parametric exposure-response family is:
\begin{equation}
      Y_{ijk} = \mu + \beta_j + \alpha_i + \nu_{ij} + \psi_{ik} +  \sum_{q=1}^Q \theta_q(s_{qij})X_{qij} + e_{ijk}.
\end{equation}
For the common case of two interventions with an optional interaction, the model is:
\begin{equation}
    Y_{ijk} = \mu + \beta_j + \alpha_i + \nu_{ij} + \psi_{ik} +  \theta_1(s_{1ij})  X_{1ij}+   \theta_2(s_{2ij}) X_{2ij} +  \theta_{1,2}(s_{1ij},s_{2ij}) X_{1ij}X_{2ij} + e_{ijk}
\end{equation} 

We start by defining a baseline \textit{immediate treatment effect} (IT) Model, which assumes instantaneous onset of the full treatment effect at the crossover time:
\begin{equation}
    f_\text{const}(s_q)=\begin{cases}
        0, & s_q=0 \\
        1 & s_q \geq 1
    \end{cases}.
\end{equation}
This is the default, non-time-varying model, used in SWDs. Below we present a selection of non-instantaneous parametric exposure-response families. 

\begin{figure}
    \centering
    \includegraphics[width=1\linewidth]{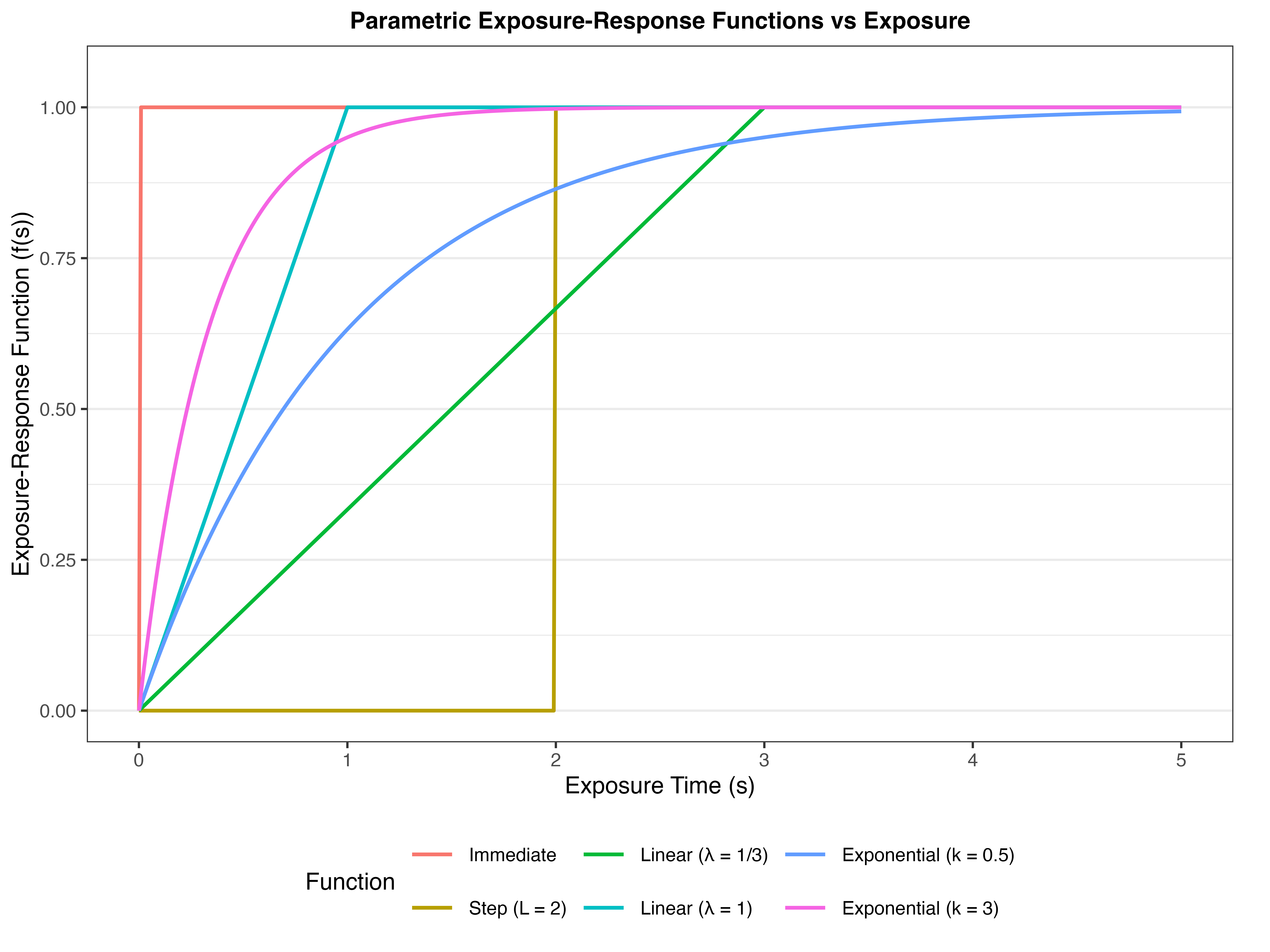}
    \caption[Parametric Exposure-Response Functions]{Parametric exposure-response functions with distinct shapes for treatment uptake over time. All functions normalized to a 5-period scenario for comparison. Each function maps exposure duration to response, with varying assumptions about latency and uptake dynamics.}
    \label{fig:Parametric}
\end{figure}

\vspace{12pt}
\noindent \textbf{Step/Threshold Effect Curve:} A simple case of a parametric-exposure response function is the \textit{step} or \textit{threshold function} defined as:
\begin{equation}
    f_\text{step}(s_q;L_q)=I(s_q \geq L_q) 
\end{equation}
where $L_q\in \mathbb{N}_0$ is the latency for intervention $q$ and $I(s_q \geq L_q)$ is the indicator function taking on value of 1 when the time since crossover to intervention $q$ is greater or equal to the latency and 0 when the time is less than the latency as shown in Figure \ref{fig:Parametric}. 

\vspace{12pt}
\noindent \textbf{Linear Effect Curve:} Another case of parametric-exposure response function is the \textit{linear function} defined as:
\begin{equation}
    f_\text{linear}(s_q;\lambda_q)=\text{min}(\lambda_q s_q,1)
\end{equation}
where $\lambda_q>0$ is the constant per-period rate of change, or ramp rate, for intervention $q$ as shown in Figure \ref{fig:Parametric}. 

\vspace{12pt}
\noindent \textbf{Exponential Effect Curve:} The parametric-exposure response \textit{exponential function} is defined as:
\begin{equation}
f_\text{exp} (s_q; k_q) = 1- e^{-k_qs_q}    
\end{equation}
where $k_q>0$ is the speed of approach to maximum realized value for intervention $q$ as shown in Figure \ref{fig:Parametric}. 

Additional parametric treatment response families, including logistic, piecewise, and power-law functions can be found in the Supplementary Materials A.

\subsection{Time-Varying Interaction Surfaces} \label{sec:timesurf} 
For $Q>1$ interventions, the assumptions of both additive treatment effects and instantaneous interaction effects may be violated and we extend to time-varying interaction effect surfaces: 
\begin{equation*}
   f_{q,q'}(s_q,s_{q'};\phi_{q,q'}), \hspace{16 pt} q \neq q'.
\end{equation*} 
For two interventions and an interaction, we define:
\begin{equation*}
    f_{1,2}(s_1, s_2; \phi_{1,2})
\end{equation*}
where $\phi_{1,2}$ is a function of a small number of interpretable parameters. The cluster-period mean model can then be written:
\begin{equation*}
    \bar{\mathbf{Y}} = \mathbf{Z}^{(\text{time})}\beta + \mathbf{Z}^{(1)}\theta_1 + \mathbf{Z}^{(2)}\theta_2 + \mathbf{Z}^{(1,2)}\theta_{1,2} + \mathbf{u}
\end{equation*}
with $\mathbf{u} \sim N(0,V)$. The form of $f_{q,q'}(s_q,s_{q'};\phi_{q,q'})$ determines the structure of $\mathbf{Z}^{(1,2)}$.

\vspace{12 pt}
\noindent \textbf{Constant Interaction:} The simplest case of a time-varying interaction surface is the \textit{constant effect} model, represented by
\begin{equation*}
    f_{1,2}(s_1,s_2; \phi_{1,2}) = 1
\end{equation*}
such that $\mathbf{Z}^{(1,2)}$ reduces to a single column $X_{1,ij}X_{2,ij}$. This model allows for an interaction term, but does not allow the effect to vary with exposure duration, similar to the immediate treatment effect. 

\begin{figure}
    \centering
    \includegraphics[width=0.9\linewidth]{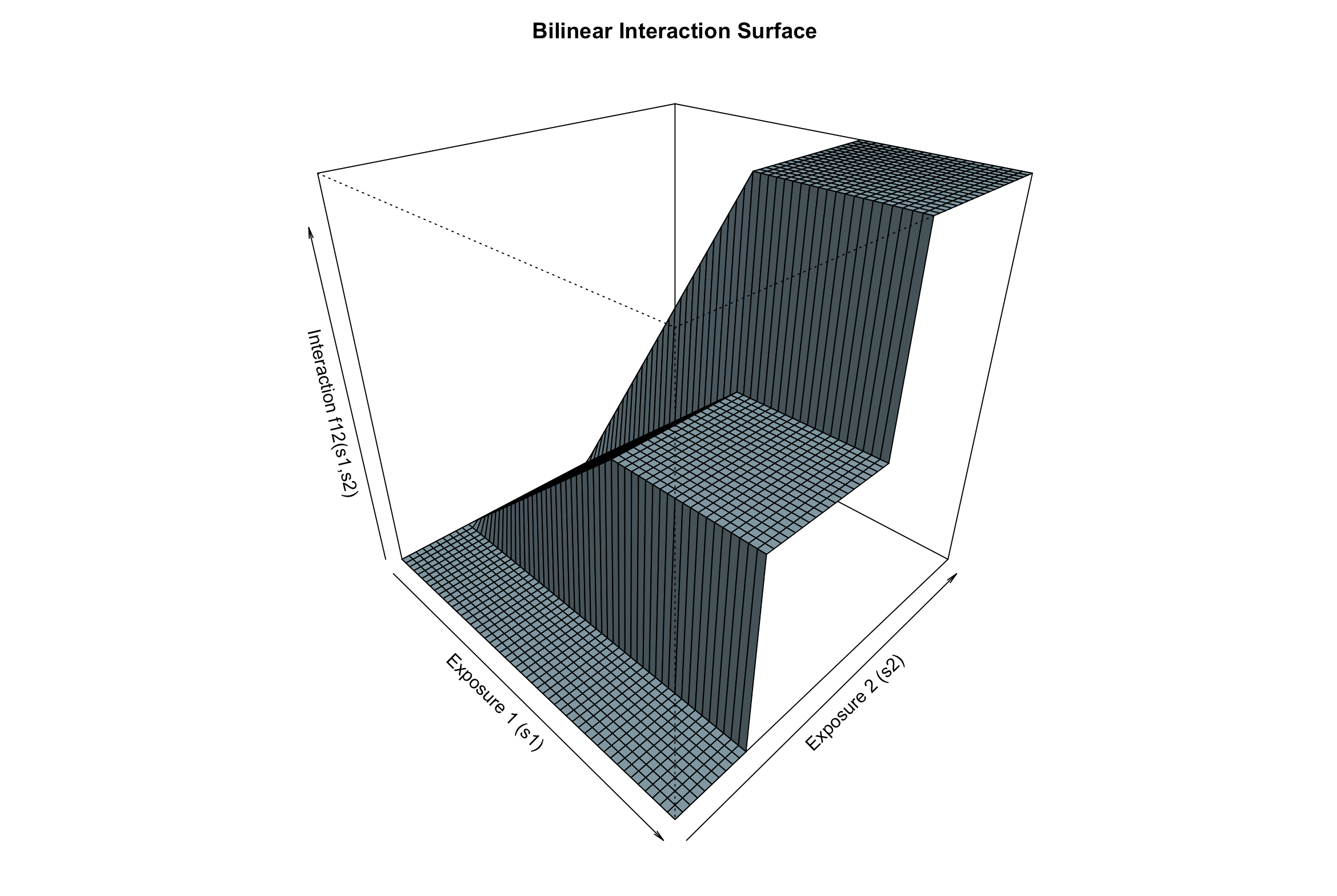}
    \caption[Example Bilinear Interaction Surface]{Example Bilinear Interaction Surface}
    \label{fig:bilinearfig}
\end{figure}

\vspace{12 pt}
\noindent \textbf{Bilinear Surface:} $f_{1,2}(s_1,s_2)=\kappa s_1 s_2$ where $\kappa>0$ is the scaling parameter. The bilinear form is an extension of the \textit{linear ramp} parametric effect curve such that the interaction grows multiplicatively with both exposures but is zero if either exposure is zero. The bilinear form grows relatively proportionally with exposure to both interventions.

Additional interaction surfaces including saturated, radial exponential, and threshold functions are described in the Supplementary Materials B.

\subsection{Power Analysis Framework for M-SWD with Flexible Correlation Structures and Expected Time-Varying Treatment Effects} \label{sec:power}

\subsubsection{Notation for Power Calculations}
We calculate power of a fixed effect using a 2-sided Wald test. Under the hypothesis, $H_0:\theta_{q}=0$ and $H_a: \theta_q=\theta_{q,a} $ power is: 
\begin{equation} \label{eq:power2}
    \text{Power} = P(|\frac{\theta_{q,a}}{\sqrt{\text{Var}(\hat{\theta_q})}}| \geq z_{1-\alpha/2} )
\end{equation}
where $\hat{\theta}_q$ is the estimated coefficient for $\theta_q$, $\alpha$ is the significance level, and $z_{1-\alpha/2}$ is the $1-\alpha/2$ percentile of the standard normal distribution.

\subsubsection{Formulation of Fixed Effects Design Matrix, $Z^*$, Under Time-Varying Treatment Effects}\label{sec:zstar}
Under delayed treatment effects, it is necessary to modify the design matrix to account for the time-varying treatment effects. We refer to this modified overall design matrix as $\mathbf{Z}^*$ and correspondingly, $\mathbf{Z}_i^*$ as the cluster-specific modified design matrix. Under this condition, the uniform integral over the interval $[j,j+1]$ provides an unbiased approximation to the average exposure effect within that period.

Under a generic parametric function for the delayed treatment effect, $f_q(s_q,\phi_q)$, we can define the cluster design matrix for cluster $i$ as:
\begin{equation*}
    \mathbf{Z}_i^* = \begin{bmatrix}
    & \mathbf{I}_{T-1} & & &  \\   \mathbf{1}_T && \text{min}\{1,\int_{s_1=s_{1ij}}^{s_{1ij}+1} f_1(s_1, \phi_1)ds_1\}  \mathbf{X}_{1i}  & \ldots & \text{min}\{1,\int_{s_Q=s_{Qij}}^{s_{Qij}+1} f_Q(s_Q, \phi_Q)ds_Q\} \mathbf{X}_{Qi} \\
   & \mathbf{0}'_{T-1} & & &
\end{bmatrix}.
\end{equation*}
Thus the period-average for intervention $q$ is 
\begin{equation*}
    Z_{ij}^{*(q)} = \int_{s_q=s_{qij}}^{s_{qij}+1} f_q(s_q,\phi_q) ds_q.
\end{equation*}

\subsubsection{Examples of $Z_i^*$}

\noindent \textbf{Step Effect Curve:} Under the step effect curve, we can calculate the $T+q$ column of the $\mathbf{Z_i^*}$ design matrix following:
\begin{align*}
  & \int_{s_q=s_{qij}}^{s_{qij}+1} f_q(s_q, \phi)ds_q =  \int_{s_q=s_{qij}}^{s_{qij}+1} \begin{cases}
        1 & s_q \geq  L_q \\
        0 & s_q < L_q 
    \end{cases} 
    =\text{max}(s_{qij}+1, L_q) - \text{max}(s_{qij},L_q)
    \end{align*}
and if all $Q$ interventions follow a step curve, we can compile the full design matrix:
\footnotesize
\begin{equation*}
    \mathbf{Z_i^*} =\begin{bmatrix}
    & \mathbf{I_{T-1}} & & &  \\   \mathbf{1_T} && [\text{max}(s_{1ij}+1, L_1) - \text{max}(s_{1ij},L_1)]\mathbf{X_{1i}}  & \ldots & [\text{max}(s_{Qij}+1, L_Q) - \text{max}(s_{Qij},L_Q)] \mathbf{X_Q}_i \\
   & \mathbf{0'_{T-1}} & & &
   \end{bmatrix}.
\end{equation*}
\normalsize

\noindent \textbf{Linear Effect Curve:} Under the linear effect curve, we can calculate the $T+q$ column of the $\mathbf{Z_i^*}$ design matrix following:
\begin{equation*}
\int_{s_q=s_{qij}}^{s_{qij}+1} f_q(s_q, \phi)ds_q =\int_{s_q=s_{qij}}^{s_{qij}+1} \lambda_q s_q \text{ }ds_q = \frac{\lambda_q s_q^2}{2}\bigg|_{s_q=s_{qij}}^{s_q=s_{qij}+1} = \frac{\lambda_q}{2}[(s_{qij}+1)^2-{s_{qij}}^2] =\lambda_q[s_{qij}+\frac{1}{2}]
\end{equation*}.

\noindent \textbf{Exponential Effect Curve:} Under the exponential effect curve, we can calculate the $T+q$ column of the $\mathbf{Z_i^*}$ design matrix following:
\begin{equation*}
\int_{s_q=s_{qij}}^{s_{qij}+1} f_q(s_q, \phi)ds_q = \int_{s_q=s_{qij}}^{s_{qij}+1} 1- e^{-k_qs_q}   \text{ } ds_q = \frac{1}{k_q}e^{-k_qs_q}+s_q\bigg|_{s_q=s_{qij}}^{s_q=s_{qij}+1} = 1-\frac{(e^{k_q}-1)e^{-(s_{qij}+1)k_q}}{k_q}
\end{equation*}.

\subsubsection{Power Calculations Under Parametric Treatment Effect Curves}\label{sec:power2}
We expand Equation \eqref{eq:power2} to estimate power under a mis-specified treatment effect curve:
\begin{equation}
    \text{Power} = P(\Big | \frac{\theta_{q,a}^{*}}{\sqrt{\text{Var}{(\hat{\theta}_{q}^*})}}\Big | \geq z_{1-\alpha/2})
\end{equation}
where $\theta_{q,a}^*$ is the implied estimand for the time varying treatment model, computed as the $T+q$ element of:
\begin{equation*}
    E[\hat{\Theta}_A^*] = (\mathbf{Z}^T\mathbf{V}^{-1}\mathbf{Z})^{-1}\mathbf{Z}^T\mathbf{V}^{-1}\mathbf{Z}^*\Theta_A 
\end{equation*}
where $\Theta_A$ is the $\Theta$ parameter vector under the alternative hypothesis and where $\text{Var}(\hat{\theta}_q^*)$ is the $[T+q, T+q]$ element of:
\begin{equation*}
    \text{Var}(\hat{\Theta}^*) = (\mathbf{Z}^{*T}\mathbf{V}^{-1}\mathbf{Z}^*)^{-1}. 
\end{equation*}

When the IT model holds, the generalized design matrix $\mathbf{Z}^*$ reduces to the standard design matrix $\mathbf{Z}$ described in equation \eqref{eq:zfull}. Thus the generalized GLS estimator and its corresponding variance under $\mathbf{Z}^*$ reduce to those derived in earlier sections and work, confirming that the standard IT model nests as a special case of the generalized M-SWD presented here. 

We provide a full worked example comparing $\mathbf{Z}^*$ and $\mathbf{Z}$ in the Supplementary materials C.

\subsection{Variance and Bias Under Time-Varying Treatment Structure}\label{sec:bias}
We can capture the anticipated information loss through the modified design matrix $\mathbf{Z}^{*}$. Under GLS, the variance-covariance matrix of the fixed-effect estimates is:
\begin{equation*}
    \mathbf{C}=(\mathbf{Z}^T\mathbf{V}^{-1}\mathbf{Z})^{-1}
\end{equation*}
while the corresponding matrix under the time-weighted design is:
\begin{equation*}
    \mathbf{C}^{*} = (\mathbf{Z}^{*T}\mathbf{V}^{-1}\mathbf{Z}^{*})^{-1}.
\end{equation*}
$L= \mathbf{C}^{*} - \mathbf{C}$ is then the efficiency loss due to the reduced information when the treatment effect is time-varying. We also calculate the bias, $E[\boldsymbol{\hat{\Theta}}]-\boldsymbol{\Theta}$, that results from estimation using the standard $\mathbf{Z}$ when the true anticipated structure follows $\mathbf{Z}^*$. Under the assumed model:

\begin{equation*}\boldsymbol{\hat{\Theta}}=(\mathbf{Z}^T\mathbf{V}^{-1}\mathbf{Z})^{-1}\mathbf{Z}^T\mathbf{V}^{-1}\mathbf{Z}\boldsymbol{\Theta} 
\end{equation*}.

If the true mean satisfies, $E[Y]=\mathbf{Z}^*\boldsymbol{\Theta}$ then
\begin{equation*}
    E[\boldsymbol{\hat{\Theta}}]=(\mathbf{Z}^T\mathbf{V}^{-1}\mathbf{Z})^{-1}\mathbf{Z}^T\mathbf{V}^{-1}\mathbf{Z^*}\boldsymbol{\Theta}. 
\end{equation*}

Thus 

\begin{equation*}
 \text{Bias}(\hat{\boldsymbol{\Theta}}) = (\mathbf{Z}^T\mathbf{V}^{-1}\mathbf{Z})^{-1}\mathbf{Z}^T\mathbf{V}^{-1}(\mathbf{Z}^*-\mathbf{Z} )\boldsymbol{\Theta}.
\end{equation*}

This result shows that when estimation proceeds under an assumed instantaneous treatment effect, the fitted coefficients no longer correspond to the marginal intervention effects but instead to weighted averages of exposure-time-specific effects.

\subsection{Treatment Effect Heterogeneity Random Effects}\label{sec:randomeffect}
An extension of the framework allows cluster-specific heterogeneity in the exposure-response function through a treatment-specific random effect:
\begin{equation*}
    \theta_q(s_{qij}) = f_q(s_{qij},\phi_q)+b_{q,i}
\end{equation*}
where $b_{q,i} \sim N(0,\tau^2_q)$. Under this framework, the covariance matrix $V$ becomes
\begin{equation}\label{eq:trtrand}
    V = \sigma^2_\nu R + \sigma^2_c I_T  + (\sigma^2_\alpha + \sigma^2_\zeta)J_T + \sum_{q=1}^Q\tau^2_qZ^{(q)}.
\end{equation}

Since this extension depends on the randomized treatment rollout pattern it generally does not yield closed-form GLS expressions. A worked example is provided in Supplementary Materials D.

\section{Analytic and Simulation Study} \label{sec:simaim2}

\subsection{Estimands Under Time-Varying Treatment Effects}
In this work, treatment effects were allowed to vary as a function of exposure time. Treatment-response functions were normalized such that the maximum realized effect ($\theta^\text{max}_q$) defined the target estimand. Under correct specification, estimators targeted this maximum effect and differences in power solely reflected information differences across exposure-response functions. In contrast, fitting an immediate-effect model targets a weighted average and therefore induces bias relative to the maximum-effect estimand. 

\subsection{Analytic and Simulation Study Setting}\label{sec:aim2resset}

\subsubsection{Designs}
\begin{figure}
    \centering
    \includegraphics[width=0.5\linewidth]{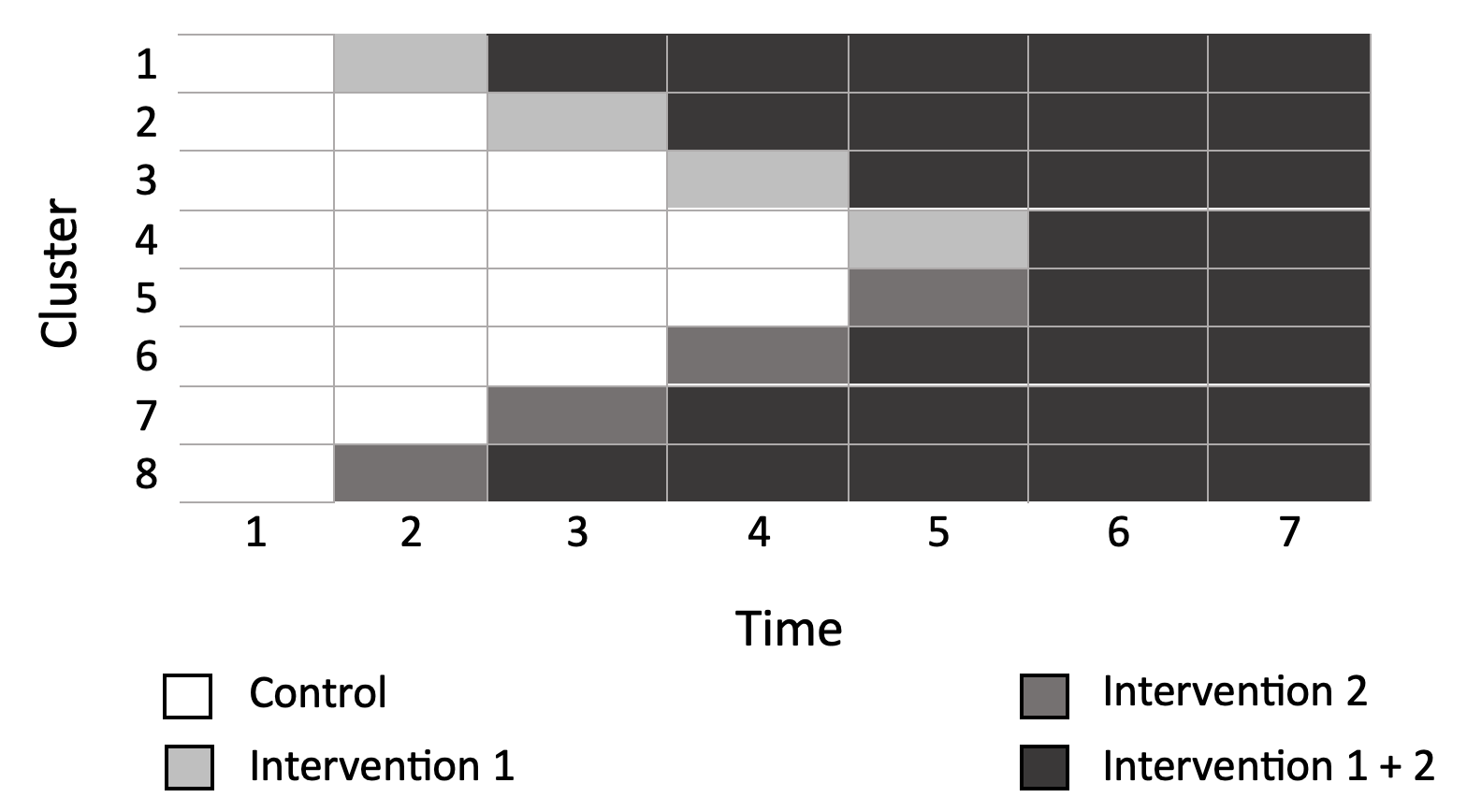}
    \caption[Aim 2 Factorial Design]{The factorial multiple intervention stepped wedge design with 8 clusters and 7 periods, as utilized throughout these results.}
    \label{fig:aim2resdes}
\end{figure}

Repeated cross-sectional designs ($\sigma^2_\psi=0$) were the primary focus as they are common in practice and isolate the impact of time-varying treatment effects. We focused on designs that were symmetric, had equal cluster sizes, and had one cluster per sequence. We assumed 50 individuals per cluster period ($n=50$), 0.2 within-period intracluster correlation (WPICC $=\rho_w=0.2$), 0.05 between-period intracluster correlation (BPICC $=\rho_b=0.05$), 2.0 total variance ($\sigma^2_y=2.0$), 1.0 Cohen's d standardized maximum treatment main effect ($\theta^{\text{max}}_{1,\text{stand}}=\theta^{\text{max}}_{2,\text{stand}}=1.0$), and 0.5 Cohen's d standardized maximum treatment interaction effect ($\theta^{\text{max}}_{\{1,2\},\text{stand}}=0.5$), unless otherwise stated. Unless otherwise stated, we utilized a compound symmetry cluster-period correlation structure parameterized by $\gamma=0.5$. We focused on complete factorial design M-SWD studies with 8 clusters ($I=8$) and 7 periods ($T=7$) as shown in Figure \ref{fig:aim2resdes}. 

\subsubsection{Evaluation Approaches}
We considered the following complementary analysis approaches:

\noindent\textbf{Method 1: Analytic GLS-Based Inference:} Closed form generalized least squares (GLS) expressions were used to compute variances, Wald test statistics, and power for treatment main effects and interactions. These calculations were based on the modified fixed-effect design matrix $\mathbf{Z}^*$ defined in Section \S\ref{sec:zstar} and the true cluster-period covariance structure. 

\noindent\textbf{Method 2: GLS Oracle Simulation:} To validate the analytic expressions and illustrate finite sample behavior, we conducted limited Monte Carlo simulations using a GLS estimator with the true known covariance structure and the correctly specified fixed effects design matrix $\mathbf{Z}^*$. 
\noindent\textbf{Method 3: Misspecified Immediate-Effect Analysis:} To evaluate the impact of analysis-stage misspecification, data generated under time-varying treatment effects were analyzed using a simplified immediate-effect model, as is typically used in practice. Bias and power were characterized analytically via the implied projection of the true mean structure onto the immediate-effect design space.   

\section{Analytic and Simulation Study Results}\label{sec:simaim2res}
\subsection{Validation of Theoretical Results}\label{sec:aim2val}

We validated the asymptotic variance and standard error expressions derived in Section \S\ref{sec:zstar} using simulations, finding essentially identical results between simulation and theoretical power (see the Supplementary Materials E).

\subsection{Power Under Properly Specified Time Varying Effects}

\subsubsection{Power Under Parametric Time Varying Treatment Effect Functions}

\begin{figure}
    \centering
    \includegraphics[width=1\linewidth]{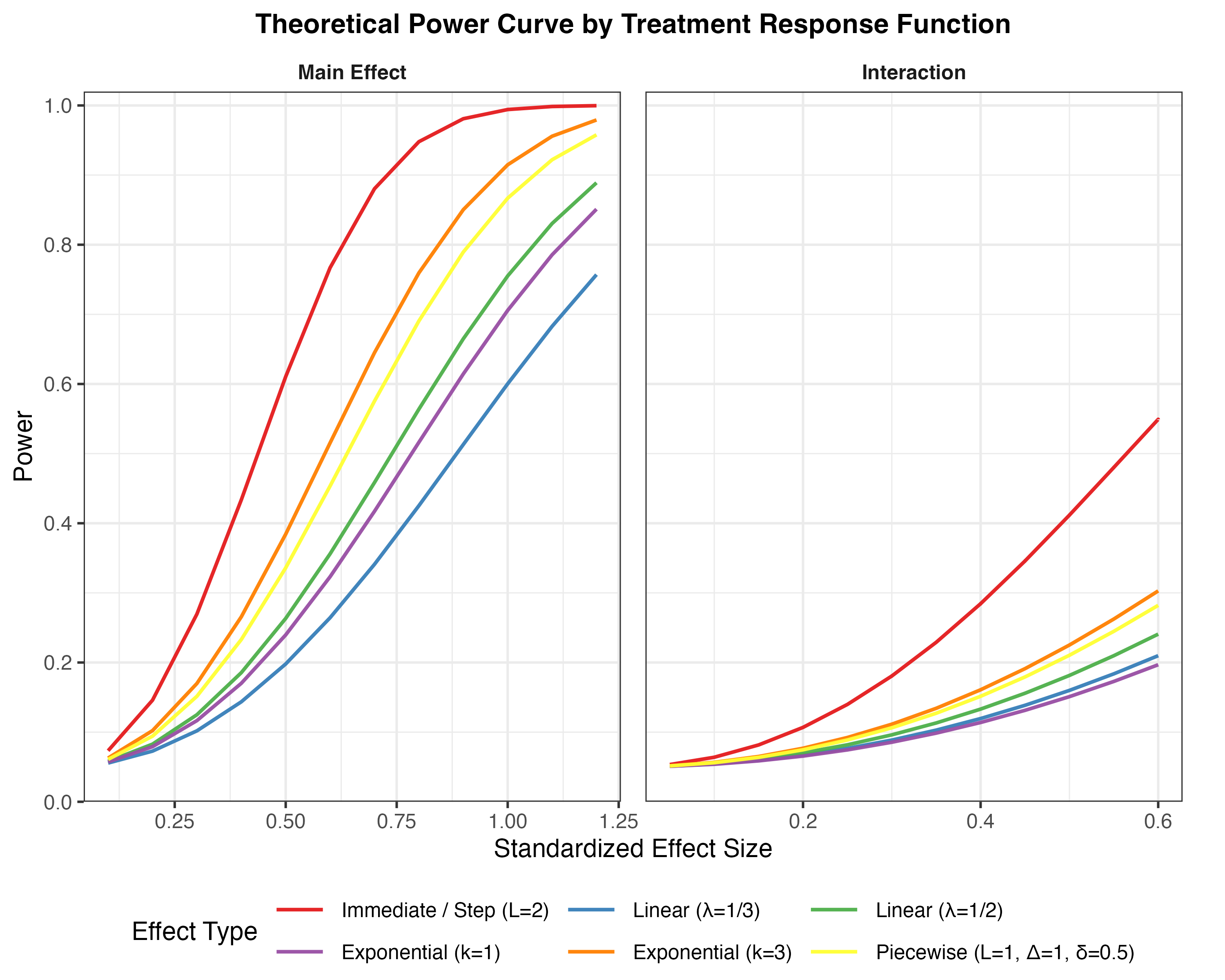}
    \caption[Theoretical Power Curves by Treatment-Response Function]{Analytic power by Cohen's d standardized effect size under various properly specified parametric time-varying treatment effect curves. Main effects shown in left panel and interaction effects shown in right panel. All results under a symmetric factorial multiple intervention SWD with $I=8$ clusters, $T=7$ periods, $n=50$ individuals per cluster-period, BPICC$=\rho_b=0.05$, WPICC$=\rho_w=0.2$, and compound-symmetry cluster-period correlation structure under moderate correlation ($\gamma=0.5$). Immediate and step (with 2 period latency) yield identical results and are collapsed to a single curve  }
    \label{fig:aim2q1effect}
\end{figure}

Figure \ref{fig:aim2q1effect} presents analytic power curves for main and interaction effects under properly specified immediate, step, linear, exponential, and piecewise time-varying treatment effect functions. Across all standardized effect sizes, immediate and step responses (with a two-period latency) yielded identical power. Under short step lag, this equivalence likely arose because the step function simply shifted the treated periods over, but did not change the total shape of the M-SWD design. In contrast, exponential, linear, and piecewise treatment-response functions reduced power, with the magnitude of the loss depending on the function shape and uptake parameter. These impacts were most substantial for interaction effects, where power remained low under all non-immediate treatment-response functions.

\subsubsection{Impact of Time-Varying Treatment Effect Function Parameters on Power}\label{sec:aim2q1_2}
\begin{figure}
    \centering
    \includegraphics[width=0.75\linewidth]{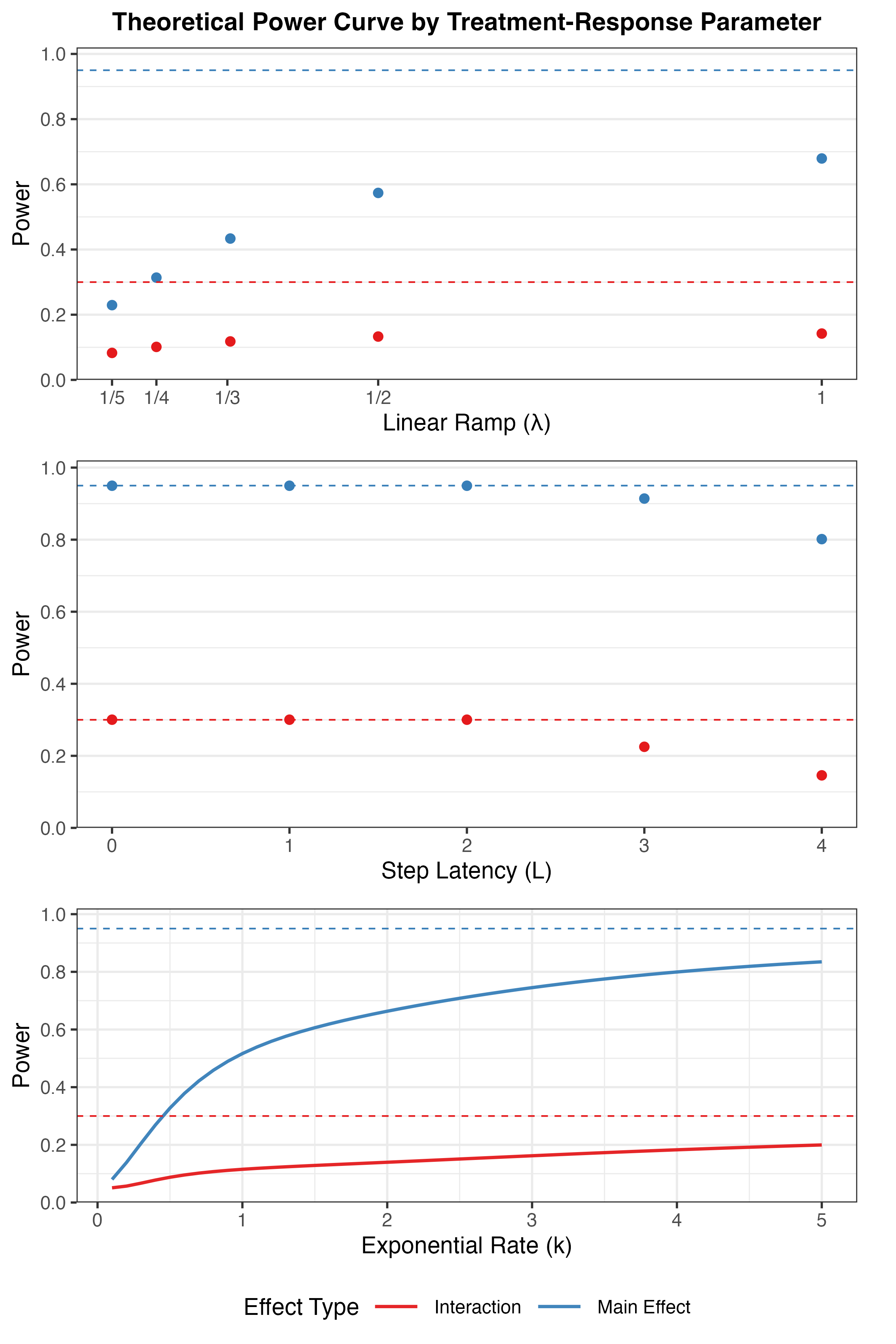}
    \caption[Theoretical Power Curves by Treatment-Response Parameter]{Theoretical power curves by time varying treatment effect function parameter ($\phi_q$). The top panel represents the impact of the linear ramp parameter ($\lambda$), middle panel step latency parameter (L), and bottom panel exponential rate (k). Blue corresponds to main effects and red to interaction effects. The blue dashed line represents the nominal main effect power for immediate treatment effect under the given parameterization (0.95) and the red dashed line the nominal interaction power for immediate treatment effect (0.30). All results calculated under an 8 cluster, 7 period factorial design with 50 individuals per cluster-period, 0.2 BPICC, 0.05 WPICC, 0.7 standardized main effect size, and 0.35 standardized interaction effect size.}
    \label{fig:aim2q1param}
\end{figure}

Figure \ref{fig:aim2q1param} presents analytic power curves for main and interaction effects under properly specified step, linear, and  exponential piecewise time-varying treatment effect functions, under various uptake parameters. Under a linear ramp treatment-response function, power increased for both main and interaction effects as the linear ramp parameter increased, but remained below the immediate treatment effect level. Similarly, under an exponential treatment-response function power increased for both main and interaction effects as the exponential rate parameter increased, approaching the level reached by the immediate effect under high rate parameters. Step latency followed a different pattern, maintaining power equivalent to immediate treatment effects under low latency ($L\leq2$ under the examined design) and then declining as latency increased.

\subsubsection{Design-Stage Cluster Requirement}
\begin{figure}
    \centering
    \includegraphics[width=0.9\linewidth]{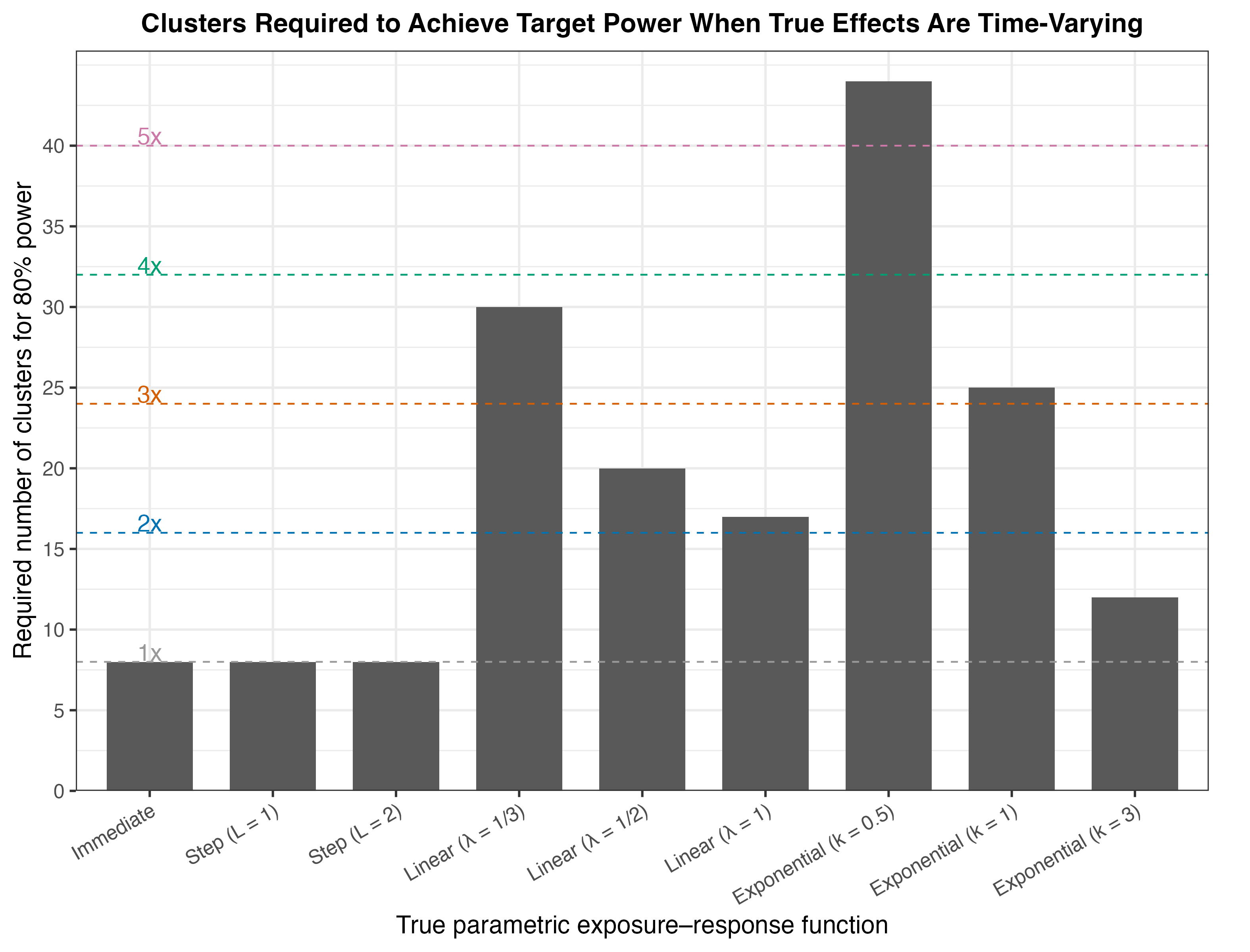}
    \caption[Clusters Required to Achieve Target Power Under Time Varying Treatment Effects]{Bar plot showing the number of clusters required to achieve 80\% power for detecting main treatment effects when the true exposure-response is time varying, but the trial is designed assuming an immediate treatment effect. All results correspond to a 7-period factorial multiple intervention stepped-wedge design with $n=50$ individuals per cluster-period, within period ICC $\rho_w=0.2$, between period ICC $\rho_b=0.05$, and standardized maximum realized main effect sizes of $d=0.48$ and maximum standardized interaction effect of $d=0.24$, based on the effect size combination to achieve 80\% power under an immediate effect model. The gray dashed horizontal line denotes the clusters required to achieve 80\% power under an immediate treatment effect model, while additional colored reference lines denote the multiplicative increases in required clusters to achieve 80\% power, relative to the baseline immediate effects design.}
    \label{fig:aim2q2d_clus}
\end{figure}

Figure \ref{fig:aim2q2d_clus} shows the number of clusters required to achieve 80\% power under each true exposure-response function. When the true treatment effect was immediate or step, the immediate-effect design assumption was appropriate and the original number of clusters was sufficient to maintain power. In contrast, when treatment effects exhibited gradual uptake, the required number of clusters was substantially higher. For slow linear and exponential uptake functions, achieving the nominal design power required approximately two to five times as many clusters as the immediate effect design.

\subsubsection{Impact of Design on Power}

To examine generalizability, we considered alternative designs as a sensitivity analysis. Under a 9 cluster concurrent design, the ordering of the power curves remained largely unchanged from those under a factorial design, with immediate and step having the highest power and linear and exponential under slow uptake having the lowest power. Overall power, however, remained lower under concurrent design, even with the additional cluster.

\begin{figure}
    \centering
    \includegraphics[width=1\linewidth]{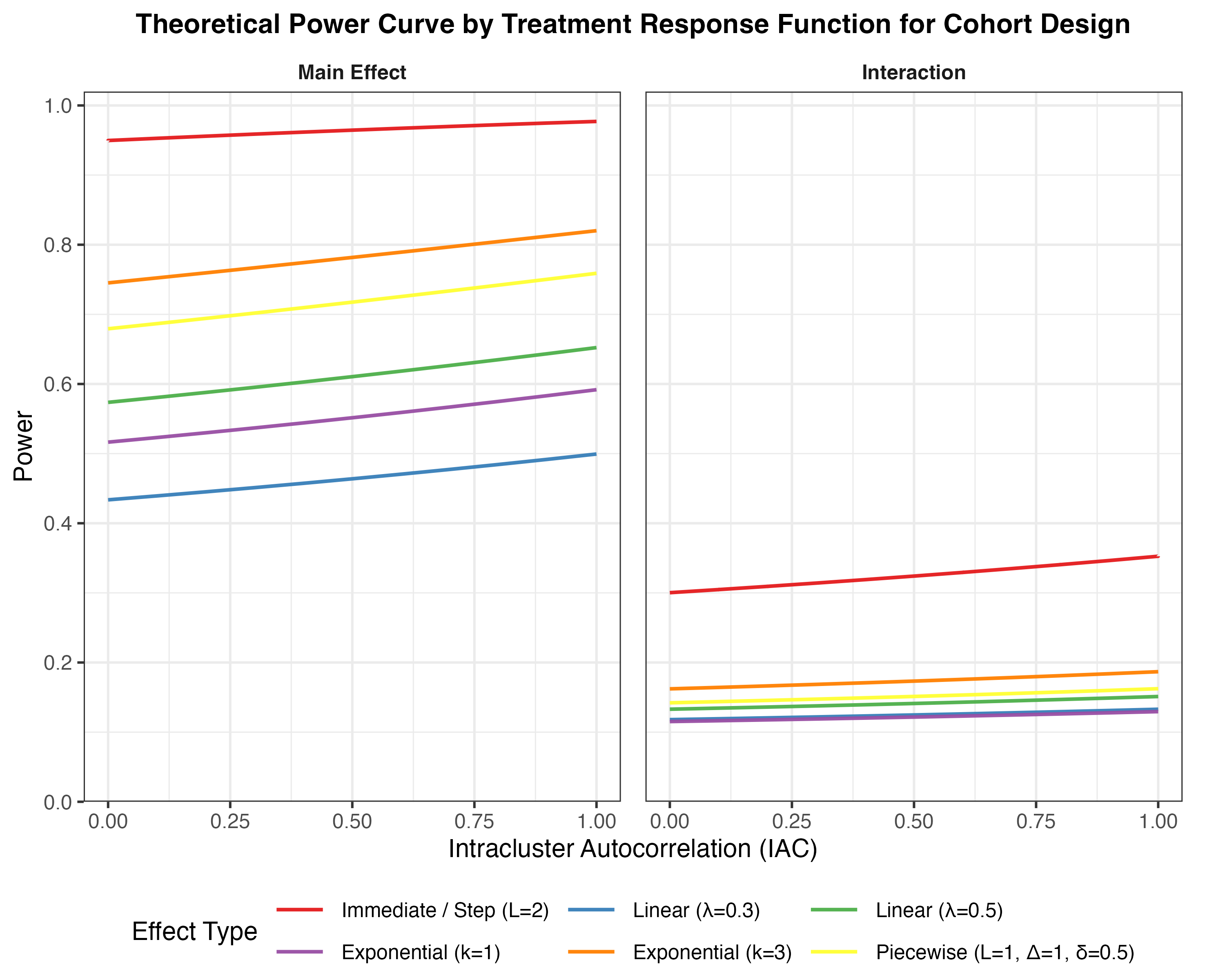}
    \caption[Power Curves Over IAC for Cohort Designs Under Parametric Time Varying Treatment Effects]{Analytic power by individual autocorrelation (IAC, $\pi$) under various properly specified parametric time-varying treatment effect curves. Main effects shown in left panel and interaction effects shown in right panel. All results under a symmetric factorial multiple intervention SWD with $I=8$ clusters, $T=7$ periods, $n=50$ individuals per cluster-period, BPICC$=\rho_b=0.05$, WPICC$=\rho_w=0.2$, and compound-symmetry cluster-period correlation structure under moderate correlation ($\gamma=0.5$). Standardized effect size of 0.7 and 0.35 used for main and and interaction effects, respectively. Immediate and step (with 2 period latency) yield identical results and are collapsed to a single curve. $\pi=0$ represents a repeated cross-sectional design. As the focus was on $\pi$, Bonferroni corrections were not utilized for this analysis.}
    \label{fig:aim2q1pi}
\end{figure}

We also examined cohort designs to test if results generalized beyond the repeated cross-sectional setting.  Figure \ref{fig:aim2q1pi} shows theoretical power as a function of the individual autocorrelation (IAC,$\pi$) under parametric time varying treatment effects. For both main and interaction effects, power increased monotonically, but modestly, with $\pi$, while the ordering of the power curves remained unchanged. Thus, the qualitative conclusions under repeated cross-sectional designs generalize to cohort designs.

\subsection{Consequences of Analysis-Stage Misspecification}

\subsubsection{Analytic Bias Under Analysis-Stage Misspecification} \label{sec:aim2biasres}

\begin{figure}
    \centering
    \includegraphics[width=0.9\linewidth]{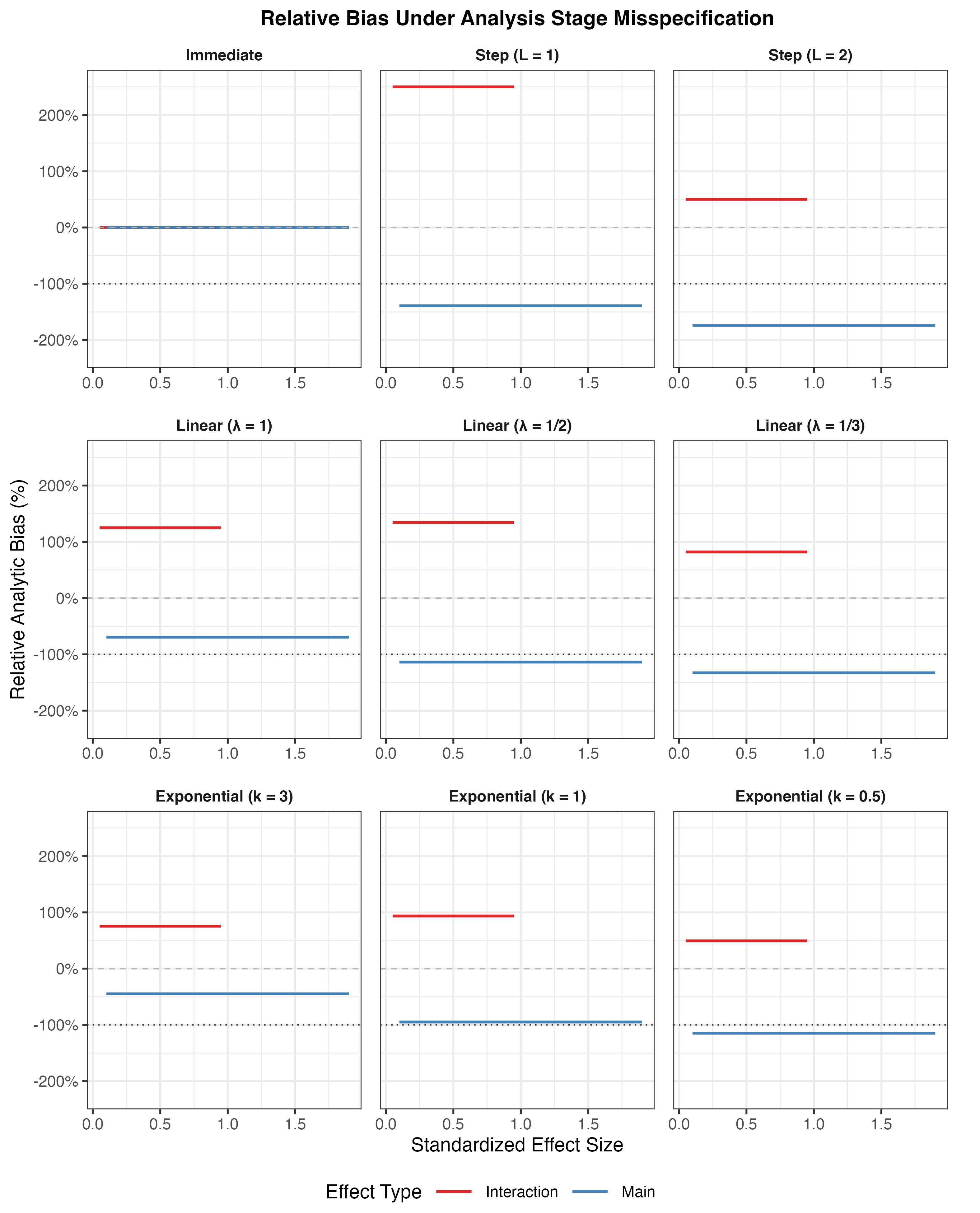}
    \caption[Analytic Bias Under Analysis-Stage Misspecification of Time Varying Treatment Effects]{Relative analytic bias percent ($(\text{E}[\hat{\theta}]-\theta)/\theta*100$) by true data generating parametric treatment response-function when an immediate effect model is assumed for analysis. All results presented under an 8 cluster, 7 period factorial M-SWD with 0.2 WPICC and 0.05 BPICC. True main effects are twice the magnitude of interaction effect across all values. The light gray dashed line at zero represents no bias and the dark gray dotted line at $-100\%$ represents the sign reversal threshold after which positive treatment effects appear negative, and vice versa.}
    \label{fig:aim2qabias}
\end{figure}

Figure \ref{fig:aim2qabias} shows the relative analytic bias induced when time-varying treatment effects were analyzed using an immediate-effect model. Bias was computed using closed-form GLS expressions derived in Section \S\ref{sec:bias}. When the immediate-effect assumption was correct, both the main and interaction estimators were unbiased, as expected. In contrast, when the true exposure-response function exhibited delayed or gradual uptake, substantial bias arose for both estimands, often exceeding 100\% relative bias. 

The direction of the bias differed systematically by effect type. Main effects were negatively biased because early post-transition periods, in which the true treatment effect had not yet reached its maximum, were coded as fully treated under the immediate-effect model. In several scenarios the relative bias for main effects was less than $-100\%$, implying sign reversal with $E[\hat{\theta}]<0$ even when $\theta>0$. In contrast, interaction effects were positively biased because the immediate-effect model overstated the magnitude of joint exposure during early jointly exposed periods. As a result, analysis-stage misspecification induced negative bias for main effects and positive bias for interaction effects. The relative bias was constant across effect sizes because it was driven by the difference between the true exposure-response design matrix and the misspecified design matrix, $\mathbf{Z}^*-\mathbf{Z}$, rather than by the strength of the treatment effect.

\paragraph{Limitations of Power Under Analysis Stage Misspecification}\label{sec:aim2powerloss}

\begin{figure}
    \centering
    \includegraphics[width=0.87\linewidth]{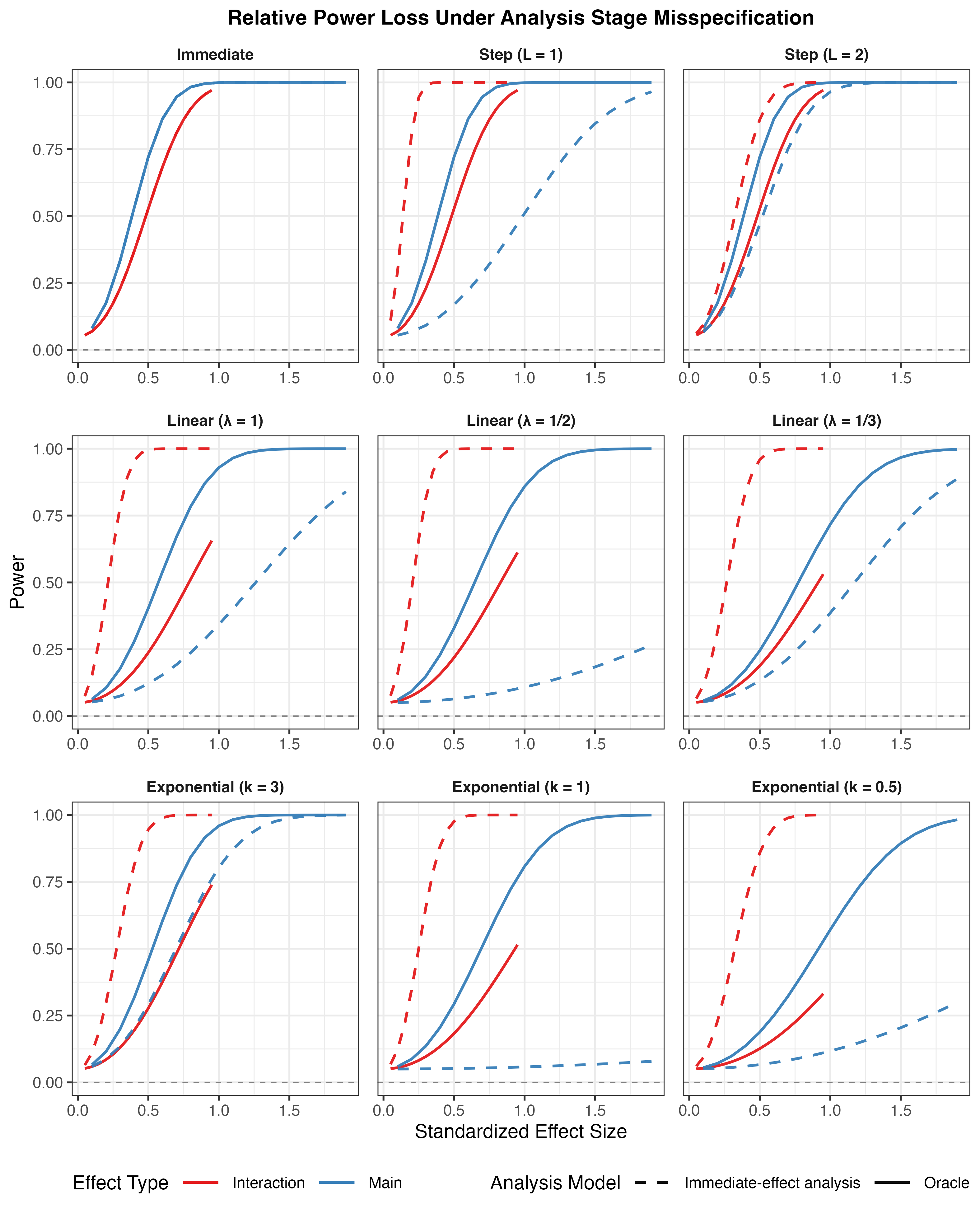}
    \caption[Analytic Power Under Analysis-Stage Misspecification of Time Varying Treatment Effects]{Analytic rejection probabilities under an oracle analysis (solid lines) and a misspecified immediate effect analysis (dashed line), by true data generating parametric treatment response-function. Red lines represent analytic power for interaction effects and blue for main effects. Results presented under an 8 cluster, 7 period factorial M-SWD with 0.2 WPICC and 0.05 BPICC. Main effects are twice the magnitude of interaction effect when the other effect varies. Differences between solid and dashed curves represent changes from estimating a bias estimand rather than true efficiency gains or losses. } 
    \label{fig:aim2q2apower}
\end{figure}

Figure \ref{fig:aim2q2apower} illustrates rejection probabilities for data generated under time-varying treatment effects but analyzed under an immediate-effect model (dashed lines) and under a properly specified oracle model (solid lines). Rejection behavior under analysis-stage misspecification was highly non-uniform, with lower rejection probabilities for main effects and higher rejection probabilities for interaction effects. These shifts do not reflect true gains or losses in efficiency, but instead arose because the misspecified analysis targeted a highly biased estimand, as shown in Figure \ref{fig:aim2qabias}. Thus, rejection probability under analysis-stage misspecification is not a valid measure of power for the true target estimand. Under analysis-stage misspecification, standard errors were generally deflated while Type I error was not impacted. This underscores that nominal Type I error alone is not sufficient to justify immediate-effect analyses when the underlying estimand is biased.

\subsection{Sensitivity to Cluster-Period Correlation Structure}
\begin{figure}
    \centering
    \includegraphics[width=1\linewidth]{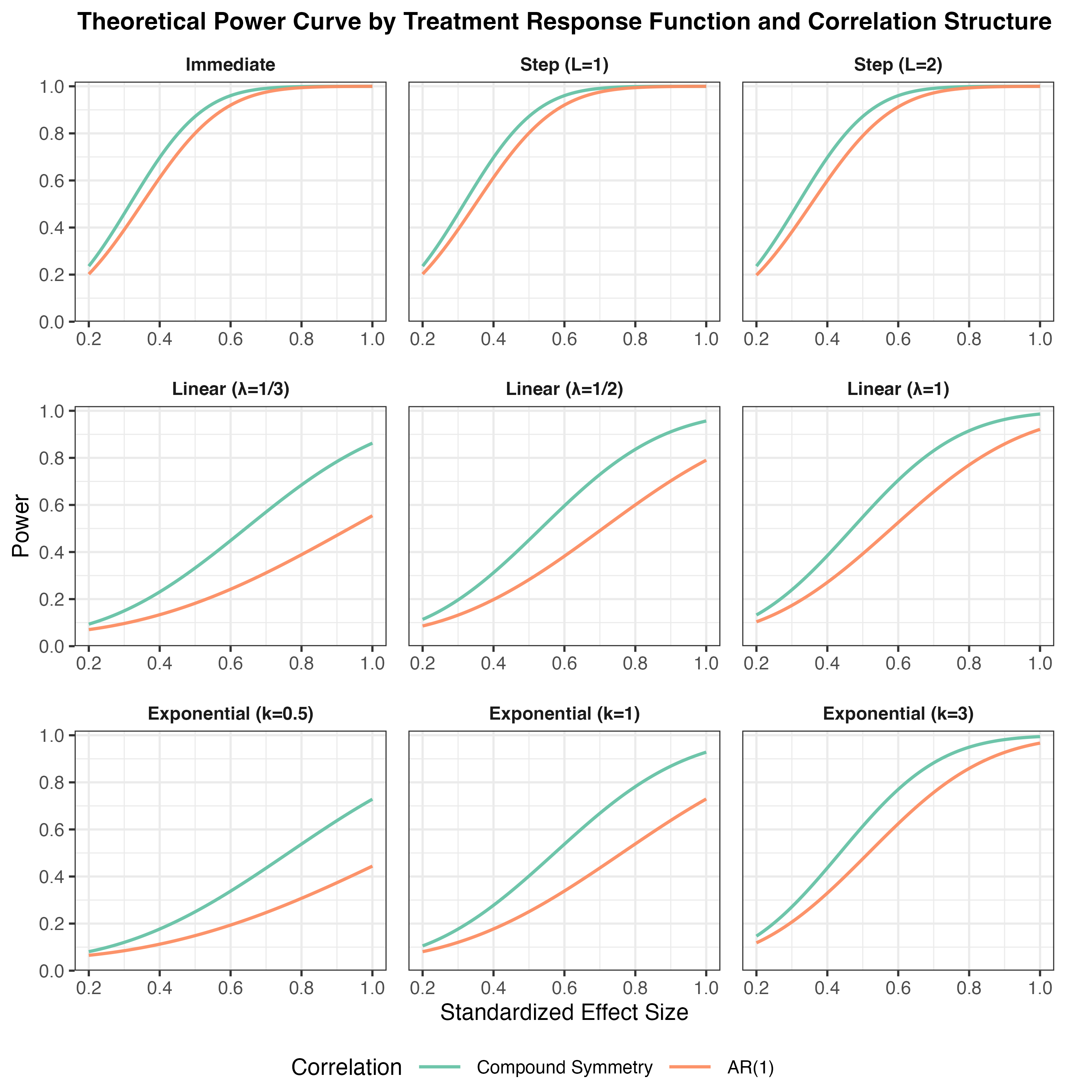}
    \caption[Theoretical Power Curves by Treatment Response Function and Correlation Structure]{Analytic power curves by treatment response function and correlation structure. Main effect shown. All results under a cross-sectional symmetric factorial multiple intervention SWD with $I=8$ cluster, $T=7$ periods, $n=50$ individuals per cluster-period, and BPICC$=\rho_b=0.05$, WPICC$=\rho_w=0.2$. Compound symmetry, represented by teal lines, is calculated under moderate-high correlation ($\gamma=0.7$) and AR(1), represented by orange lines, is calculated under moderate-high correlation ($\phi=0.7$). Bonferroni corrections were not utilized for this analysis.}
    \label{fig:aim2q3power}
\end{figure}

We next examined the sensitivity of design-stage power to assumptions about the cluster-period correlation structure. Figure \ref{fig:aim2q3power} displays theoretical power curves for main effects across a range of parametric treatment-response functions, comparing compound symmetry and AR(1) cluster-period correlation structures. Across all treatment-response functions, compound symmetry yielded uniformly higher power than AR(1). Differences between correlation structures were minimal for immediate or step functions, but became more pronounced under gradual uptake functions, particularly for slower treatment uptake. This pattern reflects the greater reliance of slower accumulating effects on later treated cluster-periods whose contributions were further reduced under temporally decaying correlation.

\subsection{Bilinear Parametric Time Varying Treatment Surfaces}
We examined a subset of results under a bilinear interaction surface to illustrate the impact of interaction uptake on model behavior. As the bilinear scaling parameter $\kappa$ decreased, power to detect interaction effects decreased monotonically, while power for main effects remained unchanged. Under a misspecified immediate-effect analysis, the magnitude and direction of the interaction bias depended strongly on the main-effect uptake shape. With immediate main effects, interaction estimates were negatively biased whereas under linear main effects they were positively biased. Supporting figures can be found in the Supplementary Materials B. These results indicate that under time-varying interaction surfaces, both power and bias for interaction effects depend on the joint uptake structure of the interaction and main effects.

\subsection{Design Robustness and Incomplete Designs}

\begin{figure}
    \centering
    \includegraphics[width=0.8\linewidth]{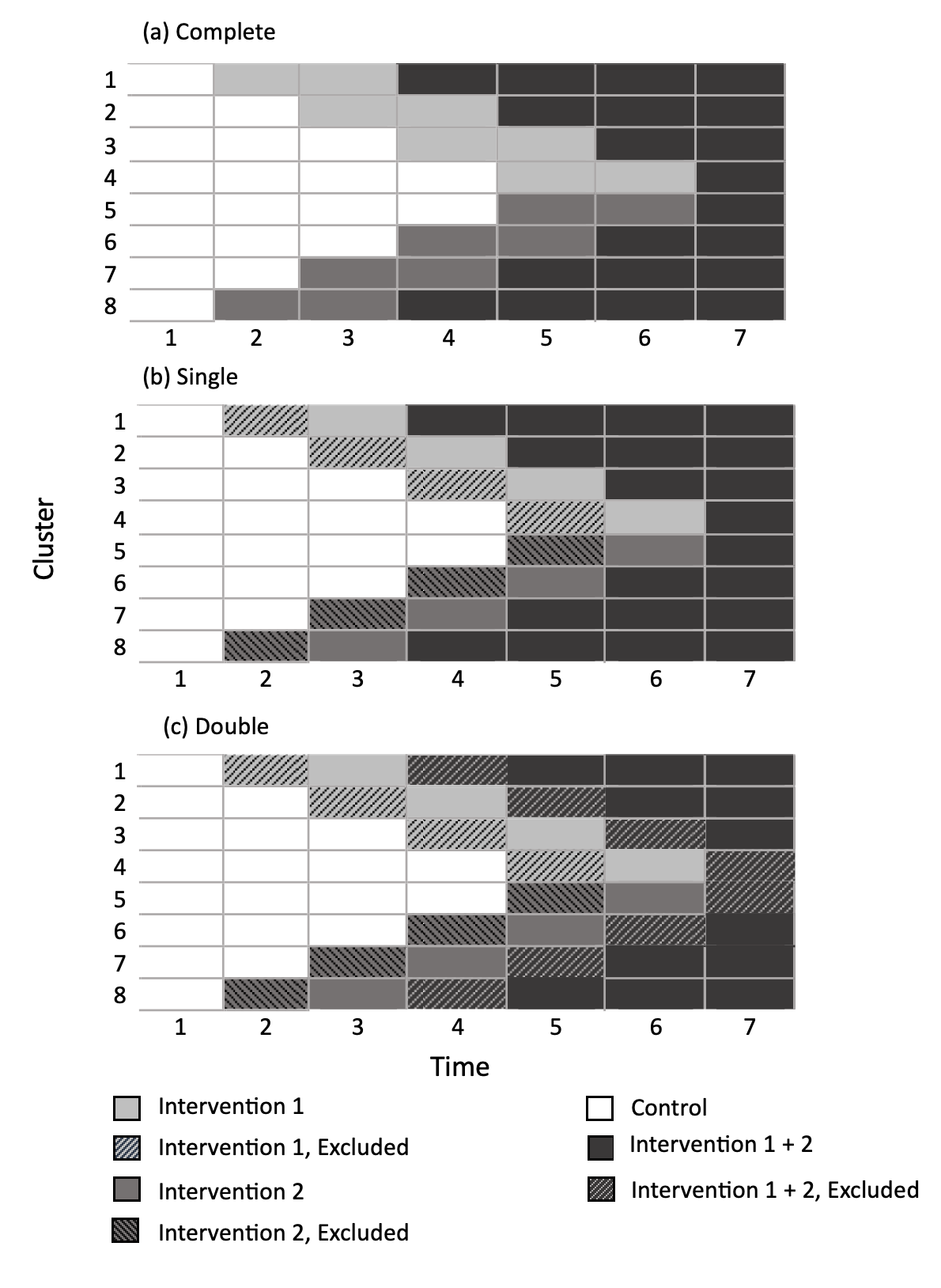}
\caption[Single and Double Incomplete Designs]{Incomplete designs for an 8 cluster, 7 period multiple intervention stepped wedge design. Solid blocks represent periods in which data are collected and analyzed and hashed blocks represent periods in which data are not collected or analyzed. Panel (a) shows a complete design with data reported and analyzed for all clusters at all periods. Panel (b) shows a single incomplete design under which data is not collected or analyzed for the first period post-transition to the \textit{first} intervention received by that cluster. Panel (c) shows a double incomplete design under which data is not collected or analyzed after the first post-transition period for \textit{each} intervention.}
    \label{fig:incompmeth}
\end{figure}

Because time-varying effects alter the distribution of information across cluster-periods, one potential design response is to exclude early post-switch observations using incomplete designs. Incomplete designs include one or more periods in which data are not collected or analyzed, such as ramp-up, training, or transition periods. We considered single-incomplete designs with one non-analyzed period after the first treatment transition for each cluster, and double-incomplete designs with one non-analyzed period after each treatment transition for each cluster. The designs considered are shown in Figure \ref{fig:incompmeth}.

\subsubsection{Power Under Incomplete Designs with Time Varying Treatment Effects}

\begin{figure}
    \centering
    \includegraphics[width=0.95\linewidth]{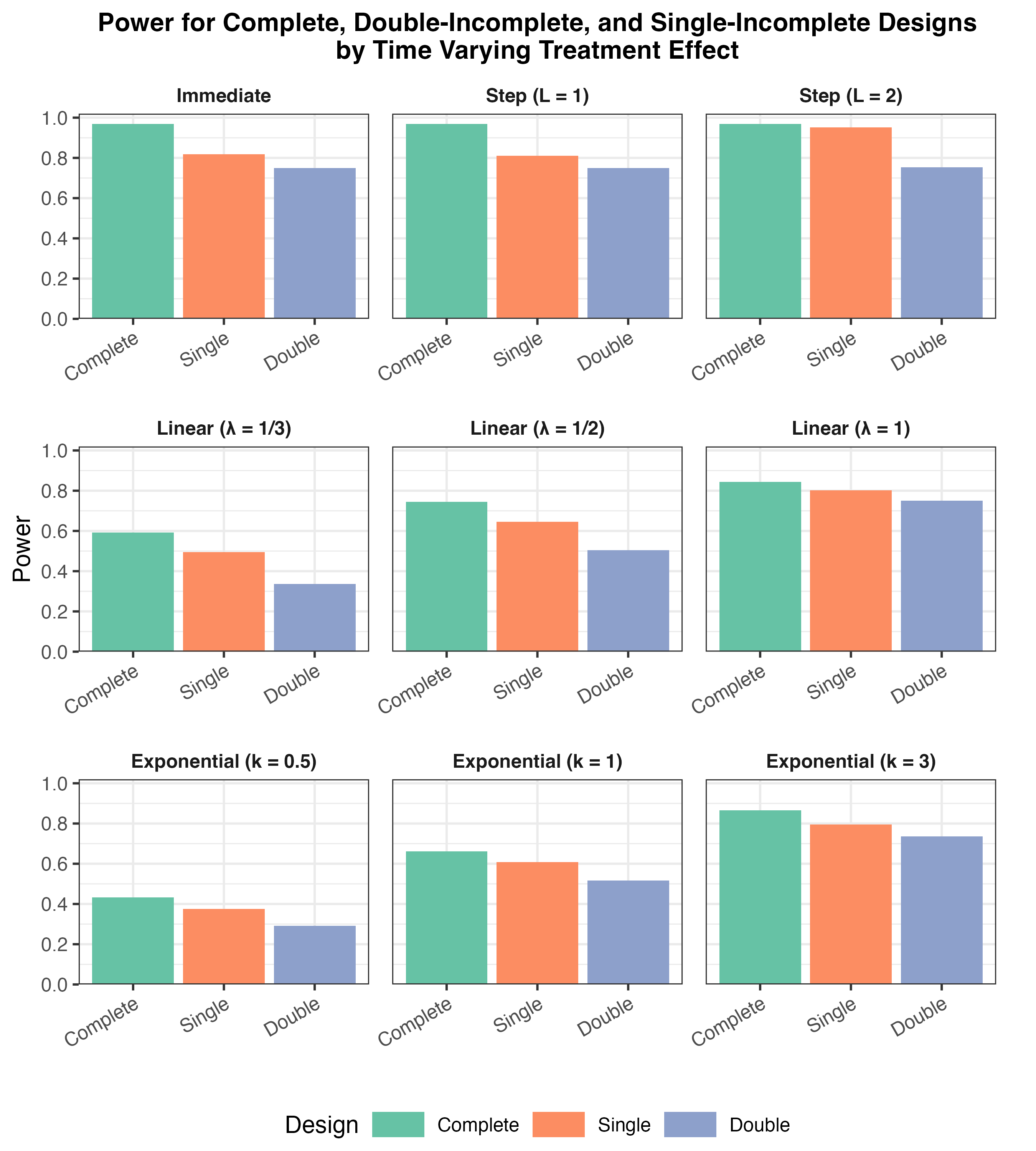}
    \caption[Power by Incomplete Design Type Under Time Varying Treatment Effects]{Power by incomplete design type and correctly specified time varying treatment effect. Bars show power under (i) complete design (green), (ii) single incomplete design that excludes the post-transition cluster-period for the first intervention received by the cluster (orange), and (iii) double-incomplete design that excludes the post-transition period for each intervention transition (blue). All results under a cross-sectional symmetric factorial multiple intervention SWD with $I=8$ cluster, $T=7$ periods, $n=50$ individuals per cluster-period, BPICC$=\rho_b=0.05$, WPICC$=\rho_w=0.2$ and standardized maximum realized main effect sizes of $d=0.7$ and maximum standardized interaction effect of $d=0.35$. Power is computed under a correctly specified time-varying treatment-response model.}
    \label{fig:aim2q4_power}
\end{figure}

Figure \ref{fig:aim2q4_power} shows power under complete, single-incomplete, and double-incomplete designs when the treatment effect was time varying. Under a properly specified time-varying analysis model, incomplete designs yielded uniformly lower power than the complete design. This was expected behavior because omitting the first post-transition period(s) removes informative post-transition observations. The magnitude of the loss was generally modest, but was larger under slower uptake functions, such as linear with small $\lambda$ or exponential with small $k$. In all scenarios, double-incomplete designs experienced greater power loss than single-incomplete because they excluded an additional post-transition period. We also examined whether extending the study duration by two periods could offset the loss of power. We found that additional periods were able to produce modest gains but did not recover the full loss. Complete results are presented in the Supplementary Materials F.

\subsubsection{Power and Bias Under Immediate Effect Analysis For Incomplete Designs}
The results above quantified the efficiency cost of incomplete designs when the analysis model was correctly specified for the true treatment-response function. In practice, however, stepped wedge trials are often analyzed under an immediate-effect assumption even when treatment effects may experience a gradual uptake or delay. We therefore examined whether excluding post-transition periods could mitigate the bias and alter rejection behavior under analysis-stage misspecification. 

\begin{figure}
    \centering
    \includegraphics[width=0.95\linewidth]{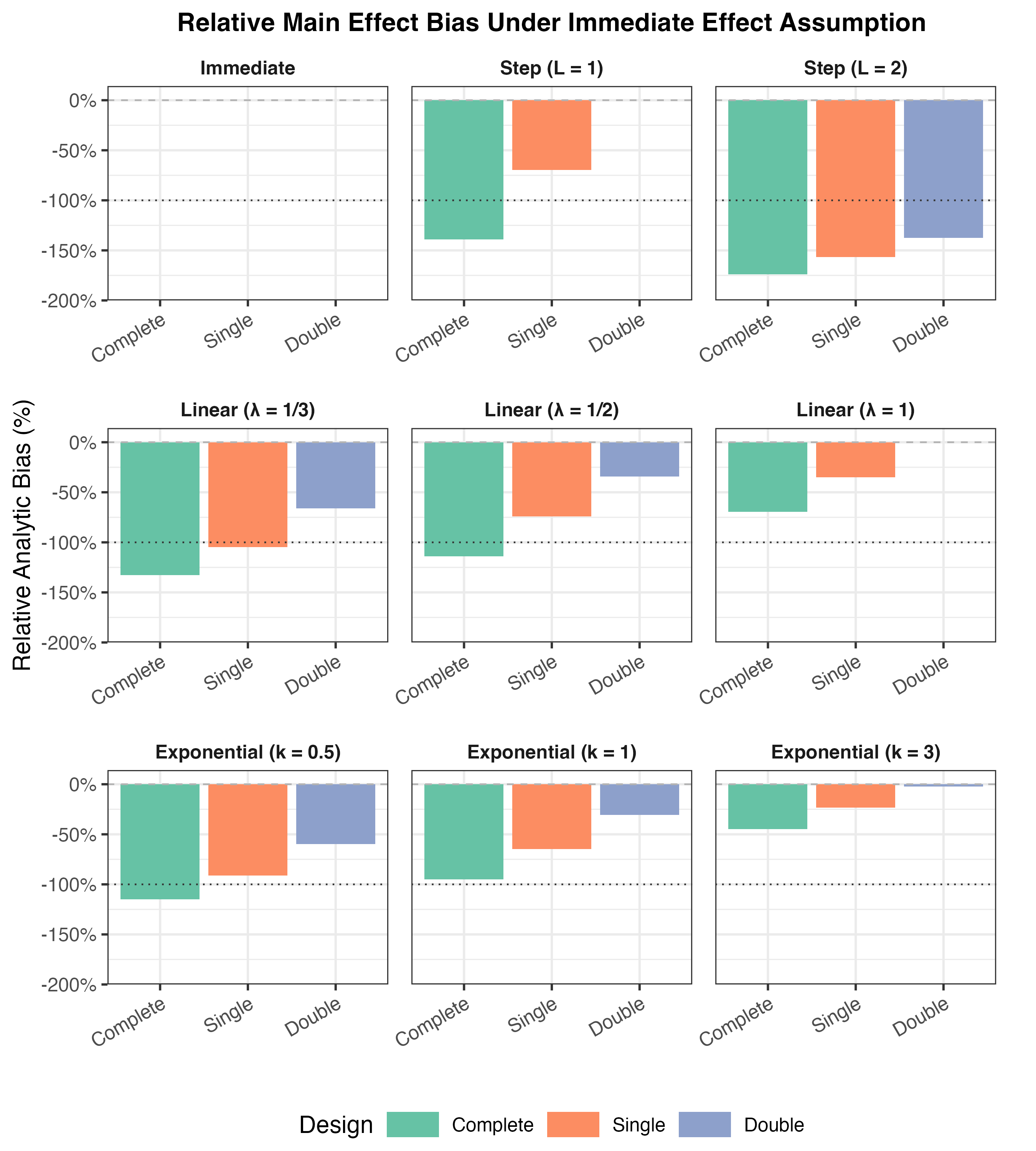}
    \caption[Relative Main Effect Bias Under Immediate Effect Assumption for Incomplete Designs]{Relative bias percent for main effects by incomplete design type and time varying treatment effect. Bias computed assuming an immediate effect analysis model, when the data is time-varying. Bars show power under (i) complete design (green), (ii) single incomplete design that excludes the post-transition cluster-period for the first intervention received by the cluster (orange), and (iii) double-incomplete design that excludes the post-transition period for each intervention transition (blue). All results under a cross-sectional symmetric factorial multiple intervention SWD with $I=8$ cluster, $T=7$ periods, $n=50$ individuals per cluster-period, BPICC$=\rho_b=0.05$, WPICC$=\rho_w=0.2$ and standardized maximum realized main effect sizes of $d=0.7$ and maximum standardized interaction effect of $d=0.35$.}
    \label{fig:aim2q4_bias_mis}
\end{figure}

Figure \ref{fig:aim2q4_bias_mis} shows the relative main effect bias under an immediate-effect analysis when the true treatment effect was time varying. Across all non-immediate treatment-response functions, main effects exhibited substantial negative bias. Incomplete designs partially mitigated this bias by excluding early post-transition periods where the discrepancy between the assumed and true effects is largest. Single-incomplete designs reduced the magnitude of bias across all scenarios, and double-incomplete designs reduced it further. However, under slower uptake functions, main effect estimators remained biased even under incomplete designs, indicating that by excluding some of the partially-treated periods, incomplete designs can improve estimator accuracy but cannot fully overcome the impact of assuming immediate effects.

\begin{figure}
    \centering
    \includegraphics[width=0.95\linewidth]{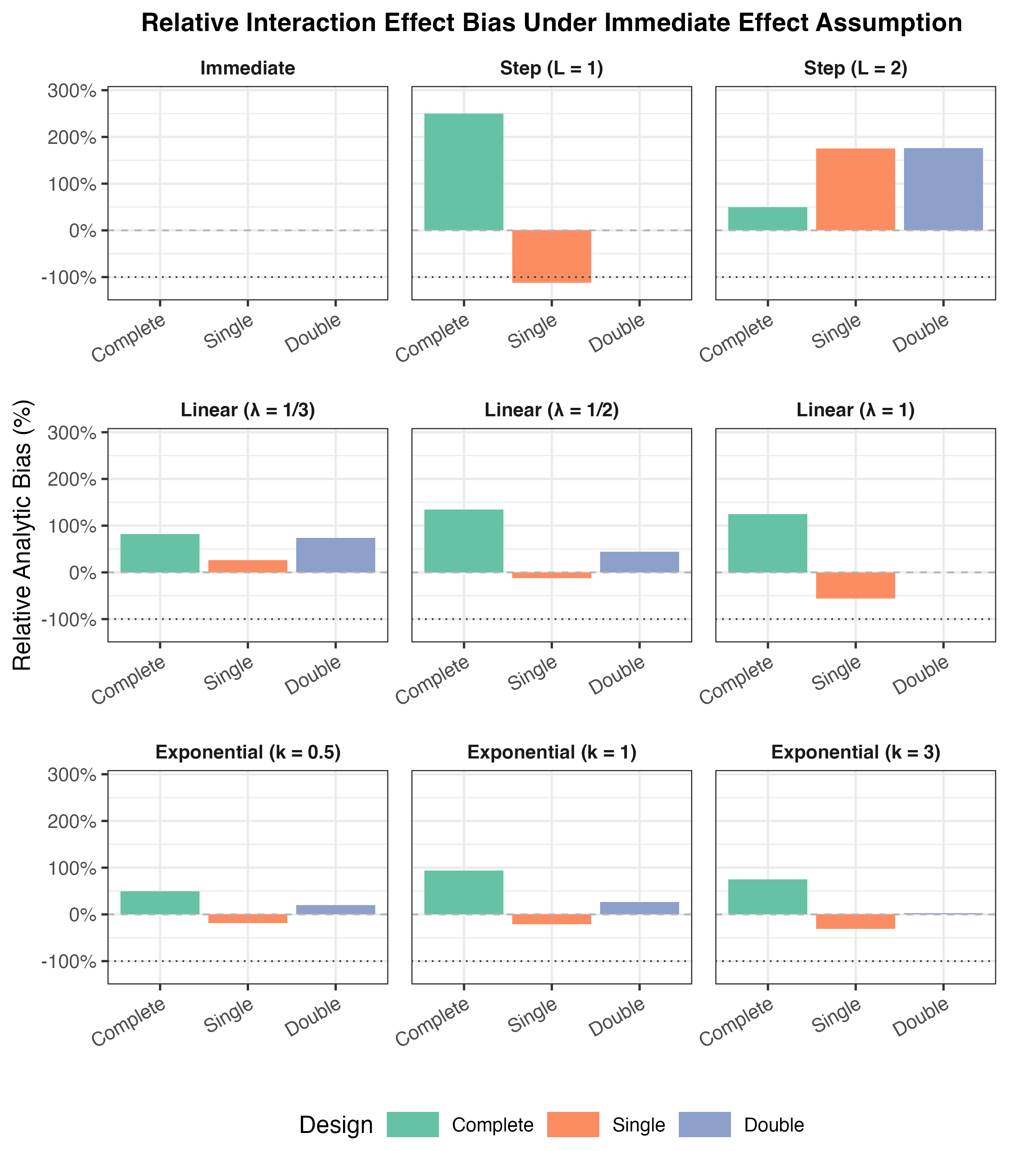}
    \caption[Relative Interaction Effect Bias Under Immediate Effect Assumption for Incomplete Designs]{Relative bias percent for interaction effects by incomplete design type and time varying treatment effect. Bias computed assuming an immediate effect analysis model, when the data is time-varying. Bars show power under (i) complete design (green), (ii) single incomplete design that excludes the post-transition cluster-period for the first intervention received by the cluster (orange), and (iii) double-incomplete design that excludes the post-transition period for each intervention transition (blue). All results under a cross-sectional symmetric factorial multiple intervention SWD with $I=8$ cluster, $T=7$ periods, $n=50$ individuals per cluster-period, BPICC$=\rho_b=0.05$, WPICC$=\rho_w=0.2$ and standardized maximum realized main effect sizes of $d=0.7$ and maximum standardized interaction effect of $d=0.35$.}
    \label{fig:aim2q4_bias_mis_int}
\end{figure}

Figure \ref{fig:aim2q4_bias_mis_int} shows that under immediate-effect analysis, interaction effects for all time varying effects were positively biased under complete designs however, incomplete designs substantially altered this behavior in non-uniform ways. For gradual uptake functions, such as linear and exponential, incomplete designs markedly reduced the bias. Single-incomplete designs often overcorrected and produced negative bias, while double-incomplete designs reduced bias closer to zero. For step function with a one-period latency ($L=1$), double-incomplete designs effectively eliminated the extreme bias seen under complete designs while single-incomplete designs overcorrected. In contrast, for for step-functions with longer latency ($L=2$), incomplete designs increased interaction bias by removing already limited information about the onset of joint exposure. Overall, incomplete designs improved interaction-effect accuracy under gradual uptake and short-latency response functions. 

\begin{figure}
    \centering
    \includegraphics[width=0.95\linewidth]{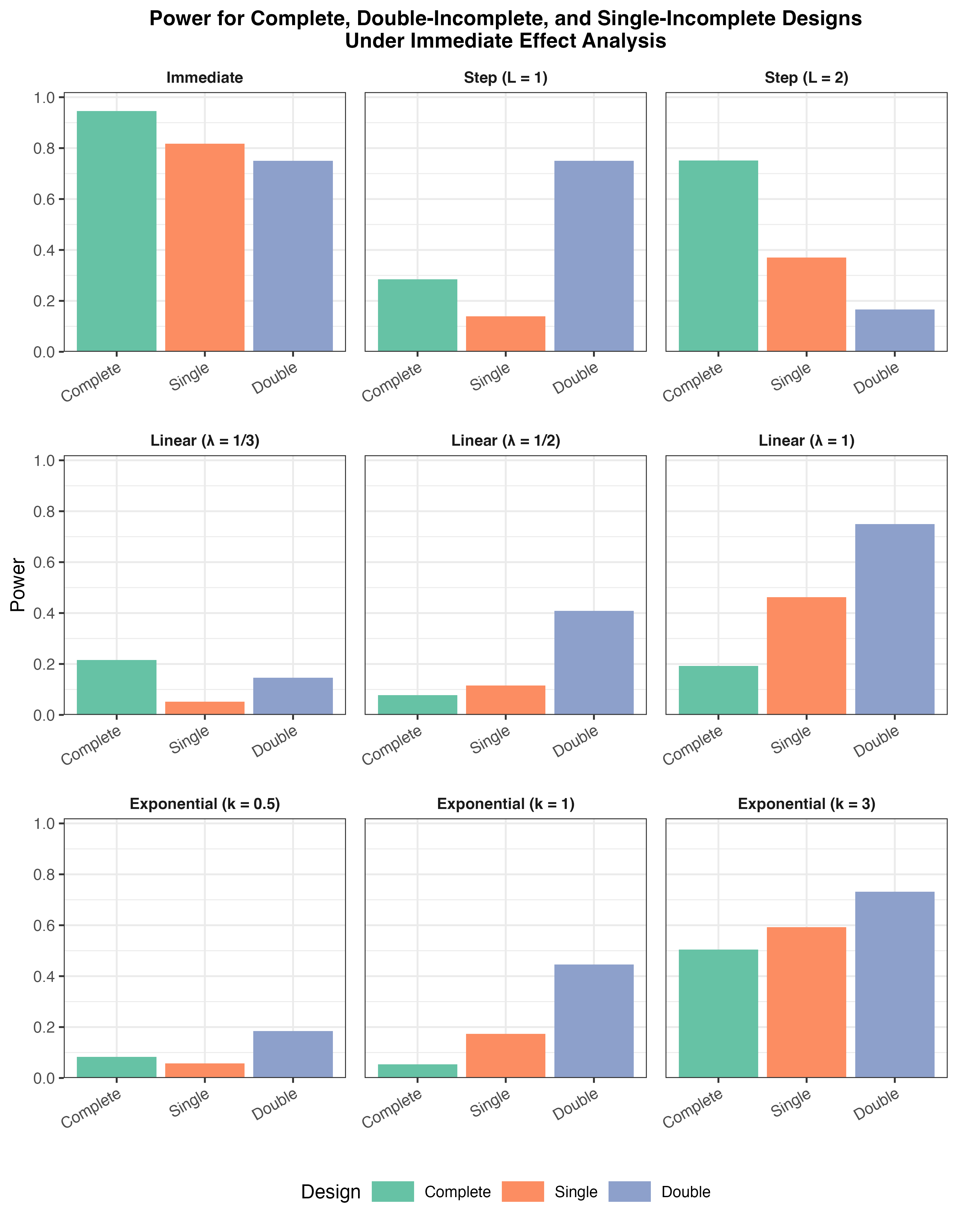}
    \caption[Power Under Immediate Effect Assumption for Incomplete Design]{Power for main effects by incomplete design type and time varying treatment effect. Power computed assuming an immediate effect analysis model, when the data is time-varying. Bars show power under (i) complete design (green), (ii) single incomplete design that excludes the post-transition cluster-period for the first intervention received by the cluster (orange), and (iii) double-incomplete design that excludes the post-transition period for each intervention transition (blue). All results under a cross-sectional symmetric factorial multiple intervention SWD with $I=8$ cluster, $T=7$ periods, $n=50$ individuals per cluster-period, BPICC$=\rho_b=0.05$, WPICC$=\rho_w=0.2$ and standardized maximum realized main effect sizes of $d=0.7$ and maximum standardized interaction effect of $d=0.35$.}
    \label{fig:aim2q4_pow_mis}
\end{figure}

Figure \ref{fig:aim2q4_pow_mis} shows main effect rejection probabilities under complete, single-incomplete, and double-incomplete designs when data generated under time-varying treatment effects were analyzed assuming an immediate treatment effect. Across all non-immediate treatment-response functions, immediate-effect misspecification led to substantial power loss for all designs. Incomplete designs alter this behavior although not uniformly. For gradual uptake functions and short step latency ($L=1$), double-incomplete designs increased rejection probability by excluding early post-transition periods under which the immediate effect assumption was most strongly violated. In contrast, for longer step latency functions ($L=2$) and slow uptake functions, incomplete designs further reduced power by excluding periods that remained important for identifying treatment onset. Single-incomplete designs were generally outperformed by either complete or double-incomplete designs. Even when incomplete designs improved rejection behavior, they did not recover the power achieved under a properly specified analysis, indicating that design stage modifications alone cannot fully compensate for analysis-stage misspecification. We note that these results should be considered in context of the bias in Figure \ref{fig:aim2q4_bias_mis}.

\section{Discussion}\label{sec:simaim2discuss}

This work investigated how time-varying treatment effects interact with multiple-intervention stepped wedge designs. We showed that treatment uptake dynamics fundamentally shape where information accrues across cluster-periods, and as such, how efficiently treatment effects can be estimated under parametric treatment-response assumptions. 

A key finding of this work is that treatment uptake dynamics centrally govern efficiency in multiple-intervention stepped wedge designs with interaction terms. When treatment effects are non-instantaneous, information is concentrated in later exposure periods, while early post-transition observations contribute relatively little signal. Main effects, which draw information from a broader set of contrasts, retained reasonable efficiency under moderate treatment-response functions. In contrast, interaction effects rely on a smaller and shorter set of cluster-period contrasts and are thus substantially more sensitive to treatment-response dynamics. As a result, interaction effects can experience pronounced power loss under gradual uptake. 

This study also distinguished the roles of design- and analysis-stage misspecification. When the analysis model correctly reflected the treatment-response structure, design-stage misspecification primarily affected efficiency rather than the target estimand itself. In contrast, analysis-stage misspecification changed the estimand being targeted. Under immediate-effect analysis, the targeted parameter was a weighted average of exposure-time specific effects rather than sustained or maximal effects, leading to systematic estimand bias relative to the intended target. Thus observed power gains under analysis-stage misspecification did not reflect increased information but rather a change in rejection behavior for biased estimands. While immediate-effect models are convenient to utilize, these results show that they can yield misleading inference when the true treatment effects evolve over time.

The evaluation of incomplete designs further illustrated the tradeoff between robustness and efficiency in the presence of parametric time-varying treatment effects. Excluding early post-transition periods in which partial or no treatment effects are realized reduces the impact of poorly aligned exposure-time contrasts under delayed uptake, but consistently reduced power under correct specification and performed poorly when early exposures carried meaningful information. These results suggest that incomplete designs are best viewed as a robustness strategy rather than a generally optimal design choice and should be considered in context with plausible treatment-response dynamics.

These findings have several implications for the design and analysis of multiple-intervention stepped wedge designs. Design- and analysis-stage decisions jointly determine how treatment effects are targeted and how efficiently they can be estimated. Based on these results, we make the following recommendations for practice:

\begin{enumerate}
    \item We recommend avoiding assumptions of instantaneous treatment effects in design-stage planning unless there is strong justification that the treatment uptake is immediate. When the true treatment effects are time varying, assuming instantaneous effects can substantially overstate efficiency, especially for interaction effects.  
    \item We recommend explicitly considering interaction effects at the design stage when they are of substantive interest. Designs that appear adequately powered for main effects can remain severely underpowered for interaction effects, especially under non-instantaneous treatment-response functions. 

    \item We recommend aligning the planned analysis model with the scientific estimand of interest. When treatment effects vary over exposure time, immediate-effect analyses target a weighted average of exposure-time specific effects rather than the sustained or maximal effect, and can therefore induce substantial bias even when Type I error appears well controlled. 

    \item We recommend considering incomplete designs as a robustness strategy when immediate-effect analyses are likely to be used. Under correctly specified time-varying analyses, excluding early post-transition periods reduces power. Under immediate effect analyses, however, excluding these periods can reduce bias and may improve rejection behavior under gradual uptake, although these gains are not uniform across uptake patterns. 

    \item We recommend considering design alternatives to M-SWDs, such as multi-arm parallel designs or crossover designs, when practical constraints allow. Although M-SWDs are adaptable under rollout and implementation constraints, non-staggered designs may be less sensitive to time-varying effects.

\end{enumerate}

While flexible analysis methods for time-varying treatment effects have been developed elsewhere, the results of this work demonstrate why relying on immediate-effect models can be misleading and clarify the design-stage consequences of these assumptions in multiple-intervention stepped wedge designs. 

These results were derived using simplified assumptions chosen to isolate the interplay between treatment uptake dynamics and design geometry in multiple-intervention stepped wedge designs. We focused on prespecified parametric treatment-response functions under symmetric factorial designs with balanced cluster-period sample sizes and assumed correctly specified correlation structures to separate treatment-uptake effects from correlation misspecification. Repeated cross-sectional designs were emphasized to avoid additional complexity from individual-level autocorrelation. These assumptions allowed for clear interpretation of how treatment-response dynamics redistribute information across cluster-periods and impact efficiency and bias, particularly for interaction effects. Future work could extend this framework to asymmetric designs, alternative interaction surfaces or random treatment effect heterogeneity, unequal cluster sizes, non-normal outcomes, and settings with correlation structure misspecification.


\bibliographystyle{unsrtnat}
\bibliography{References}

@article{sundin_power_2022,
	title = {Power analysis for stepped wedge trials with multiple interventions},
	volume = {41},
	issn = {0277-6715, 1097-0258},
	url = {https://onlinelibrary.wiley.com/doi/10.1002/sim.9301},
	doi = {10.1002/sim.9301},
	language = {en},
	number = {8},
	urldate = {2023-10-03},
	journal = {Statistics in Medicine},
	author = {Sundin, Phillip and Crespi, Catherine M.},
	month = apr,
	year = {2022},
	pages = {1498--1512},
}

@article{lyons_proposed_2017,
	title = {Proposed variations of the stepped-wedge design can be used to accommodate multiple interventions},
	volume = {86},
	issn = {1878-5921},
	doi = {10.1016/j.jclinepi.2017.04.004},
	language = {eng},
	journal = {Journal of Clinical Epidemiology},
	author = {Lyons, Vivian H. and Li, Lingyu and Hughes, James P. and Rowhani-Rahbar, Ali},
	month = jun,
	year = {2017},
	pmid = {28412466},
	pmcid = {PMC5835387},
	pages = {160--167},
}

@article{hussey_design_2007,
	title = {Design and analysis of stepped wedge cluster randomized trials},
	volume = {28},
	issn = {1551-7144},
	doi = {10.1016/j.cct.2006.05.007},
	language = {eng},
	number = {2},
	journal = {Contemporary Clinical Trials},
	author = {Hussey, Michael A. and Hughes, James P.},
	month = feb,
	year = {2007},
	pmid = {16829207},
	pages = {182--191},
}

@article{finney_rutten_evaluating_2018,
	title = {Evaluating the impact of multilevel evidence-based implementation strategies to enhance provider recommendation on human papillomavirus vaccination rates among an empaneled primary care patient population: a study protocol for a stepped-wedge cluster randomized trial},
	volume = {13},
	issn = {1748-5908},
	shorttitle = {Evaluating the impact of multilevel evidence-based implementation strategies to enhance provider recommendation on human papillomavirus vaccination rates among an empaneled primary care patient population},
	url = {https://doi.org/10.1186/s13012-018-0778-x},
	doi = {10.1186/s13012-018-0778-x},
	number = {1},
	urldate = {2024-02-17},
	journal = {Implementation Science},
	author = {Finney Rutten, Lila J. and Radecki Breitkopf, Carmen and St. Sauver, Jennifer L. and Croghan, Ivana T. and Jacobson, Debra J. and Wilson, Patrick M. and Herrin, Jeph and Jacobson, Robert M.},
	month = jul,
	year = {2018},
	pages = {96},
}

@article{pol_effectiveness_2017,
	title = {Effectiveness of sensor monitoring in an occupational therapy rehabilitation program for older individuals after hip fracture, the {SO}-{HIP} trial: study protocol of a three-arm stepped wedge cluster randomized trial},
	volume = {17},
	issn = {1472-6963},
	shorttitle = {Effectiveness of sensor monitoring in an occupational therapy rehabilitation program for older individuals after hip fracture, the {SO}-{HIP} trial},
	url = {https://doi.org/10.1186/s12913-016-1934-0},
	doi = {10.1186/s12913-016-1934-0},
	language = {en},
	number = {1},
	urldate = {2024-02-17},
	journal = {BMC Health Services Research},
	author = {Pol, Margriet C. and ter Riet, Gerben and van Hartingsveldt, Margo and Kröse, Ben and de Rooij, Sophia E. and Buurman, Bianca M.},
	month = jan,
	year = {2017},
	pages = {3},
}

@article{zhu_enhancing_2023,
	title = {Enhancing doctor-patient relationships in community health care institutions: the {Patient} {Oriented} {Four} {Habits} {Model} ({POFHM}) trial—a stepped wedge cluster randomized trial protocol},
	volume = {23},
	issn = {1471-244X},
	shorttitle = {Enhancing doctor-patient relationships in community health care institutions},
	url = {https://www.ncbi.nlm.nih.gov/pmc/articles/PMC10308620/},
	doi = {10.1186/s12888-023-04948-w},
	urldate = {2024-02-17},
	journal = {BMC Psychiatry},
	author = {Zhu, Yunying and Li, Sisi and Zhang, Ruotong and Bao, Lei and Zhang, Jin and Xiao, Xiaohua and Jiang, Dongdong and Chen, Wenxiao and Hu, Chenying and Zou, Changli and Zhang, Jingna and Zhu, Yong and Wang, Jianqiu and Liang, Jinchun and Yang, Qian},
	month = jun,
	year = {2023},
	pmid = {37380993},
	pmcid = {PMC10308620},
	pages = {476},
}

@article{pol_effectiveness_2019,
	title = {Effectiveness of sensor monitoring in a rehabilitation programme for older patients after hip fracture: a three-arm stepped wedge randomised trial},
	volume = {48},
	issn = {0002-0729},
	shorttitle = {Effectiveness of sensor monitoring in a rehabilitation programme for older patients after hip fracture},
	url = {https://doi.org/10.1093/ageing/afz074},
	doi = {10.1093/ageing/afz074},
	number = {5},
	urldate = {2024-02-20},
	journal = {Age and Ageing},
	author = {Pol, Margriet C and ter Riet, Gerben and van Hartingsveldt, Margo and Kröse, Ben and Buurman, Bianca M},
	month = sep,
	year = {2019},
	pages = {650--657},
}

@article{piszczek_stepped-wedge_2015,
	title = {Stepped-wedge trial design to evaluate {Ebola} treatments},
	volume = {15},
	issn = {1473-3099, 1474-4457},
	url = {https://www.thelancet.com/journals/laninf/article/PIIS1473-3099(15)00078-X/fulltext},
	doi = {10.1016/S1473-3099(15)00078-X},
	language = {English},
	number = {7},
	urldate = {2024-02-22},
	journal = {The Lancet Infectious Diseases},
	author = {Piszczek, Jolanta and Partlow, Eric},
	month = jul,
	year = {2015},
	pmid = {26122441},
	note = {Publisher: Elsevier},
	pages = {762--763},
}

@article{cook_statistical_2016,
	title = {Statistical {Lessons} {Learned} for {Designing} {Cluster} {Randomized} {Pragmatic} {Clinical} {Trials} from the {NIH} {Health} {Care} {Systems} {Collaboratory} {Biostatistics} and {Design} {Core}},
	volume = {13},
	issn = {1740-7745},
	url = {https://www.ncbi.nlm.nih.gov/pmc/articles/PMC5025337/},
	doi = {10.1177/1740774516646578},
	number = {5},
	urldate = {2024-02-22},
	journal = {Clinical trials (London, England)},
	author = {Cook, Andrea J and Delong, Elizabeth and Murray, David M and Vollmer, William M and Heagerty, Patrick J},
	month = oct,
	year = {2016},
	pmid = {27179253},
	pmcid = {PMC5025337},
	pages = {504--512},
}

@article{federico_ethical_2022,
	title = {Ethical and epistemic issues in the design and conduct of pragmatic stepped-wedge cluster randomized clinical trials},
	volume = {115},
	issn = {1551-7144},
	url = {https://www.ncbi.nlm.nih.gov/pmc/articles/PMC9272561/},
	doi = {10.1016/j.cct.2022.106703},
	urldate = {2024-02-23},
	journal = {Contemporary clinical trials},
	author = {Federico, Carole A. and Heagerty, Patrick J. and Lantos, John and O’Rourke, Pearl and Rahimzadeh, Vasiliki and Sugarman, Jeremy and Weinfurt, Kevin and Wendler, David and Wilfond, Benjamin S. and Magnus, David},
	month = apr,
	year = {2022},
	pmid = {35176501},
	pmcid = {PMC9272561},
	pages = {106703},
}

@article{brown_stepped_2006,
	title = {The stepped wedge trial design: a systematic review},
	volume = {6},
	issn = {1471-2288},
	shorttitle = {The stepped wedge trial design},
	url = {https://www.ncbi.nlm.nih.gov/pmc/articles/PMC1636652/},
	doi = {10.1186/1471-2288-6-54},
	urldate = {2024-02-27},
	journal = {BMC Medical Research Methodology},
	author = {Brown, Celia A and Lilford, Richard J},
	month = nov,
	year = {2006},
	pmid = {17092344},
	pmcid = {PMC1636652},
	pages = {54},
}

@article{kasza_impact_2019,
	title = {Impact of non-uniform correlation structure on sample size and power in multiple-period cluster randomised trials},
	volume = {28},
	issn = {0962-2802},
	url = {https://doi.org/10.1177/0962280217734981},
	doi = {10.1177/0962280217734981},
	language = {en},
	number = {3},
	urldate = {2024-03-22},
	journal = {Statistical Methods in Medical Research},
	author = {Kasza, J and Hemming, K and Hooper, R and Matthews, JNS and Forbes, AB},
	month = mar,
	year = {2019},
	note = {Publisher: SAGE Publications Ltd STM},
	pages = {703--716},
}

@article{thompson_bias_2017,
	title = {Bias and inference from misspecified mixed-effect models in stepped wedge trial analysis},
	volume = {36},
	copyright = {© 2017 The Authors. Statistics in Medicine published by John Wiley \& Sons Ltd.},
	issn = {1097-0258},
	url = {https://onlinelibrary.wiley.com/doi/abs/10.1002/sim.7348},
	doi = {10.1002/sim.7348},
	language = {en},
	number = {23},
	urldate = {2024-03-22},
	journal = {Statistics in Medicine},
	author = {Thompson, Jennifer A. and Fielding, Katherine L. and Davey, Calum and Aiken, Alexander M. and Hargreaves, James R. and Hayes, Richard J.},
	year = {2017},
	note = {\_eprint: https://onlinelibrary.wiley.com/doi/pdf/10.1002/sim.7348},
	pages = {3670--3682},
}

@article{kasza_inference_2019,
	title = {Inference for the treatment effect in multiple-period cluster randomised trials when random effect correlation structure is misspecified},
	volume = {28},
	issn = {0962-2802},
	url = {https://doi.org/10.1177/0962280218797151},
	doi = {10.1177/0962280218797151},
	language = {en},
	number = {10-11},
	urldate = {2024-03-22},
	journal = {Statistical Methods in Medical Research},
	author = {Kasza, Jessica and Forbes, Andrew B},
	month = nov,
	year = {2019},
	note = {Publisher: SAGE Publications Ltd STM},
	pages = {3112--3122},
}

@article{kenny_2022,
	title = {Analysis of stepped wedge cluster randomized trials in the presence of a time-varying treatment effect},
	issn = {0277-6715},
	url = {https://www.ncbi.nlm.nih.gov/pmc/articles/PMC9481733/},
	doi = {10.1002/sim.9511},
	urldate = {2025-03-19},
	journal = {Statistics in medicine},
	author = {Kenny, Avi and Voldal, Emily and Xia, Fan and Heagerty, Patrick J. and Hughes, James P.},
	month = jun,
	year = {2022},
	pmid = {35774016},
	pmcid = {PMC9481733},
	pages = {10.1002/sim.9511},
}

@article{nickless_2018,
	title = {Mixed effects approach to the analysis of the stepped wedge cluster randomised trial—{Investigating} the confounding effect of time through simulation},
	volume = {13},
	issn = {1932-6203},
	url = {https://www.ncbi.nlm.nih.gov/pmc/articles/PMC6292598/},
	doi = {10.1371/journal.pone.0208876},
	number = {12},
	urldate = {2025-03-19},
	journal = {PLoS ONE},
	author = {Nickless, Alecia and Voysey, Merryn and Geddes, John and Yu, Ly-Mee and Fanshawe, Thomas R.},
	month = dec,
	year = {2018},
	pmid = {30543671},
	pmcid = {PMC6292598},
	pages = {e0208876},
}

@article{chen_timevar_2025,
      title={Time-varying treatment effect models in stepped-wedge cluster-randomized trials with multiple interventions}, 
      author={Zhe Chen and Wei Wang and Yingying Lu and Scott D. Halpern and Katherine R. Courtright and Fan Li and Michael O. Harhay},
      year={2025},
      eprint={2504.14109},
      archivePrefix={arXiv},
journal={Arxiv},
      url={https://arxiv.org/abs/2504.14109}, 
}

@article{hughes_current_2015,
	series = {10th {Anniversary} {Special} {Issue}},
	title = {Current issues in the design and analysis of stepped wedge trials},
	volume = {45},
	issn = {1551-7144},
	url = {https://www.sciencedirect.com/science/article/pii/S1551714415300434},
	doi = {10.1016/j.cct.2015.07.006},
	urldate = {2025-10-10},
	journal = {Contemporary Clinical Trials},
	author = {Hughes, James P. and Granston, Tanya S. and Heagerty, Patrick J.},
	month = nov,
	year = {2015},
	pages = {55--60},
}

@misc{paper1arxiv,
      title={A generalized multiple-intervention stepped wedge design framework for treatment effect estimation in the presence of non-uniform cluster-period correlation structures}, 
      author={Samantha M. Levy and Jose-Miguel Yamal},
      year={2026},
      eprint={2606.23499},
      archivePrefix={arXiv},
      primaryClass={stat.ME},
      url={https://arxiv.org/abs/2606.23499}, 
}

\pagebreak

\section{Supplemental Materials}

\subsection{Supplemental Materials A: Additional Parametric Exposure-Response Functions}\label{sect:addparam}

\vspace{12pt}
\noindent \textbf{Logistic Effect Curve:} The \textit{logistic} or \textit{sigmoidal function} is defined as:
\begin{equation}
    f(s_q; \gamma_q, m_q) = \frac{1}{1 + \text{exp}[-\gamma_q(s_q-m_q)]}
\end{equation}
where $\gamma_q >0$ is the steepness of the curve for intervention $q$ and $m_q \geq 0$ is the midpoint of the curve for intervention $q$. 

\vspace{12pt}
\noindent \textbf{Piecewise Effect Curves:} The \textit{piecewise function} is defined as: 
\begin{equation}
    f_\text{piece}(s_q;L_q,\delta_q, \Delta_q) = \begin{cases}
0, & s_q < L_q,\\[6pt]
\delta_q, & L_q \leq s_q \leq L_q + \Delta_q,\\[6pt]
1, & s_q > L_q + \Delta_q,
\end{cases}
\end{equation}
where $L_q \in \{0,1,2, \ldots \}$ is the latency for intervention $q$, $\delta_q$ is the plateau height for intervention $q$, and $\Delta_q$ is the plateau duration for intervention $q$. 

We note that $\delta_q$ could also be its own function such that $\delta_q=\delta_q(s_q) \in [0,1]$. The most likely scenario is a linear ramp such that:
\begin{equation}
    f_\text{piece-ramp}(s_q;L_q,\delta_q(.), \Delta_q) = \begin{cases}
0, & s_q < L_q,\\[6pt]
\delta_q(s_q)=\lambda_q(s_q-L_q), & L_q \leq s_q \leq L_q + \Delta_q,\\[6pt]
1, & s_q > L_q + \Delta_q,
\end{cases}
\end{equation}
where to maintain continuity at $s_q=L_q+\Delta_q$, $\Delta_q=1/\lambda_q$. 

\vspace{12pt}
\noindent \textbf{Power Law Effect Curve:} The \textit{power law function} is defined as:
\begin{equation}
    f_\text{power}(s_q;\alpha_q)=\text{min}\{s_q^{\alpha_q},1\}
\end{equation}
where $\alpha_q \in (0,1]$ is the concavity. 

\subsection{Supplemental Materials B: Additional Parametric Interaction Surfaces And Their Results}

\paragraph{Saturated Surface:}
A \textit{saturated surface} allows the interaction effect to vary freely across both exposure durations, represented by:
\begin{equation*}
    f_{1,2}(s_1,s_2) = I_{\{(s_1,s_2)\}},
\end{equation*}
where $I_{\{(s_1,s_2)\}}$ denotes an indicator vector taking the value $1$ for the corresponding joint exposure-time combination $(s_1,s_2)$ and $0$ otherwise. Thus the block $\mathbf{Z}^{(1,2)}$ contains indicator columns for every possible exposure-time combination. This follows a non-parametric structure similar to that presented by Kenny et al. \cite{kenny_2022} and Chen et al. \cite{chen_timevar_2025} but expanded to M-SWD interaction surfaces. This model places no structural assumptions on the interaction surface, allowing for full heterogeneity in the interaction, however it also introduces $(T-1)^2$ parameters which can lead to significant identifiability concerns. 

\begin{figure}
    \centering
    \includegraphics[width=1.1\linewidth]{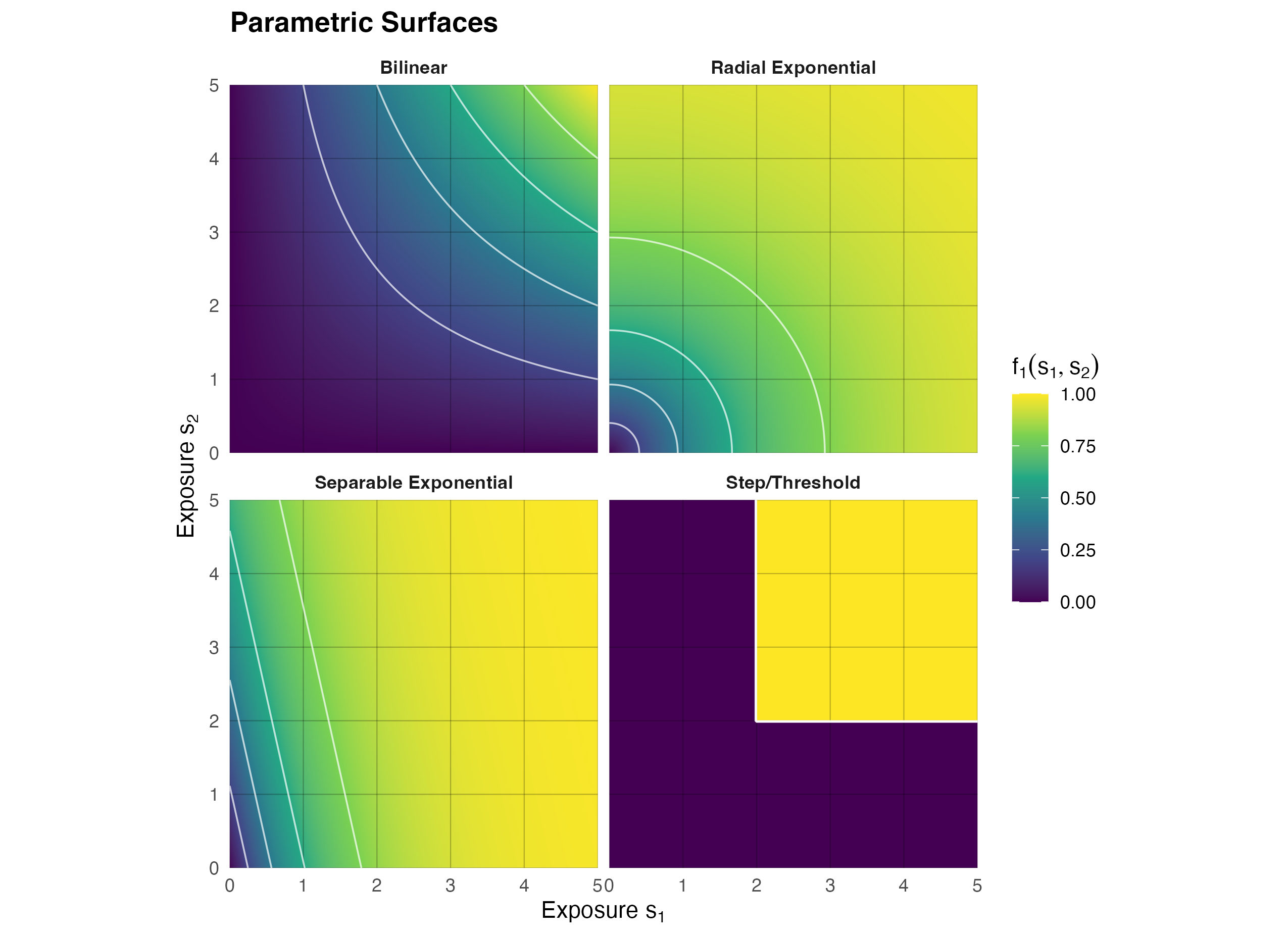}
    \caption[Parametric Exposure Response Interaction Surfaces]{Continuous representation of parametric surfaces for the interaction effect as a function of exposure times $s_1$ and $s_2$. The top left panel demonstrates a \textit{bilinear} surface ($\kappa s_1 s_2$) with $\kappa=1/25$, the top right panel demonstrates a \textit{radial exponential} surface ($\tau \left(1-\exp[-\xi\sqrt{s_1^2+s_2^2}]\right)$) with $\tau = 1, \xi=0.55$, the bottom left panel demonstrates a \textit{separable exponential} surface ($\tau\left(1-\exp[-\xi_1 s_1 - \xi_2 s_2]\right)$) with $\tau=1, \xi_1=0.9, \xi_2=0.2$, and the bottom right panel demonstrates a \textit{step/threshold} surface ($\delta I[s_1\geq L_1,\ s_2 \geq L_2]$) with $L_1=2, L_2=2, \delta=1$. Color intensity represents the strength of the interaction effect $f_{1,2}(s_1,s_2)$, faint grid lines section interval exposure periods, and white contour lines represent equal effect values.}
    \label{fig:surface}
\end{figure}

\vspace{12pt}
\noindent \textbf{Exponential Surface:} $f_{1,2}(s_1,s_2; \tau, \xi_1, \xi_2) = \tau(1-\text{exp}[-\xi_1 s_1 - \xi_2 s_2])$ where $\tau>0$ is the maximum joint effect and $\xi_q>0$ is the rate at which exposure to intervention $q$ contributes to the interaction effect. Under this surface, the interaction grows quickly at first and then plateaus, with the plateau height being set by $\tau$ and the rates being set by $\xi_1$ and $\xi_2$ as shown in Figure \ref{fig:surface}. 

\vspace{12pt}
\noindent \textbf{Radial Exponential Surface:} $f_{1,2}(s_1,s_2;\tau,\xi) = \tau (1-\text{exp}[-\xi\sqrt{(s_1^2+s_2^2)}])$ where $\tau>0$ is the maximum joint effect and $\xi>0$ controls the rate at which the interaction accumulates with total combined exposure duration. Unlike the separable exponential surface, the radial form depends only on the magnitude of joint exposure $r=\sqrt{s_1^2+s_2^2}$ and is therefore symmetric in $s_1$ and $s_2$. As $\xi$ goes to 0, the interaction increases slowly with exposure, whereas larger values of $\xi$ produce more rapid growth toward the asymptote $\tau$. This formulation yields smooth, concentric contour lines corresponding to equal levels of combined exposure intensity, as shown in Figure \ref{fig:surface}.


\vspace{12 pt}
\noindent \textbf{Step/Threshold Interaction Surface:} $f_{1,2}(s_1,s_2; \delta , L_1, L_2) = \delta^*I[s_1\geq L_1, s_2 \geq L_2]$ where $\delta$ is the incremental effect once both interventions have been in place long enough and $L_1 \text{ and } L_2$ are the corresponding minimum exposure times for $\delta$ to be initiated. The step/threshold function shows a fixed latency before the interaction appears once both interventions have been active for a set number of periods as shown in Figure \ref{fig:surface}. Under this surface, the effects are assumed to be additive until a given post-implementation duration after which a synergistic effect is observed. 

\begin{figure}
    \centering
    \includegraphics[width=1\linewidth]{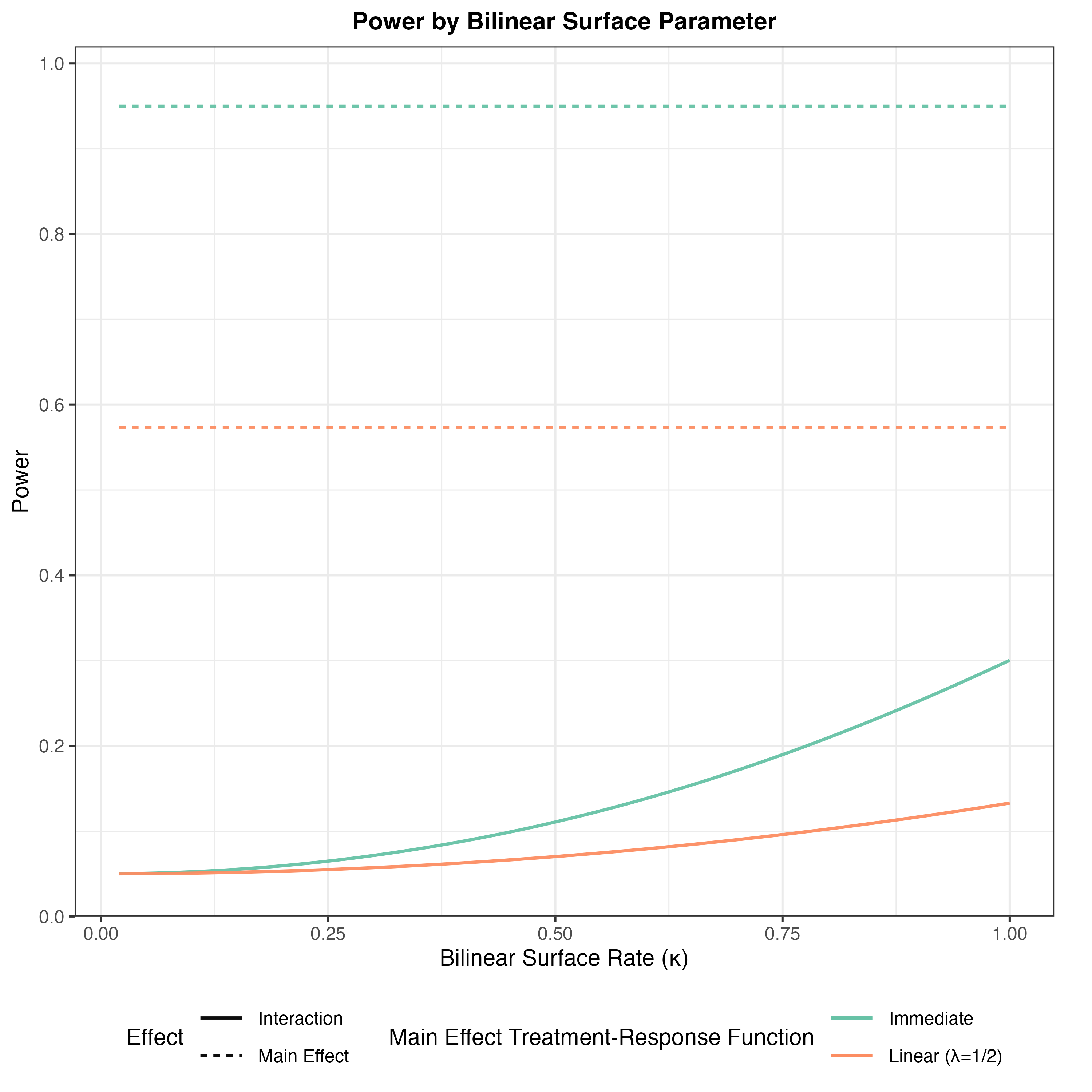}
    \caption[Power by Bilinear Surface Parameter]{Analytic power curves by bilinear surface rate parameter ($\kappa$). Dashed line represent main effects and solid lines interaction effects. Green lines represent power computed under an immediate effect assumption and orange linear ($\lambda=1/2$) treatment-response function for main effects. All results under a cross-sectional symmetric factorial multiple intervention SWD with compound symmetry ($\gamma=0.5$), $I=8$ cluster, $T=7$ periods, $n=50$ individuals per cluster-period, BPICC$=\rho_b=0.05$, WPICC$=\rho_w=0.2$, standardized main effect $d=0.7$, and standardized interaction effect $d=0.35$. Bonferroni corrections were not utilized for this analysis.}
    \label{fig:aim2_q5_kappa_power}
\end{figure}

\vspace{12 pt}
\noindent \textbf{Results Under Bilinear Parametric Treatment Surface:}
We examined a subset of results under a bilinear interaction surface to illustrate the impact of interaction uptake on model behavior. Figure \ref{fig:aim2_q5_kappa_power} presents theoretical power for detecting interaction effects as a function of the bilinear surface scaling parameter $\kappa$, under both constant and linear ($\lambda=1/2$) main effect response functions under proper model specification. As shown by the dashed lines, power to detect main effects remained constant under changing bilinear scaling parameter. In contrast, power to detect interaction effects decreased monotonically as $\kappa$ approached zero, reflecting slower accumulation of interaction effects over time. Under an immediate main effect, power to detect interaction effects remained uniformly higher than under a moderate linear effect. Thus we found that under bilinear interaction surfaces, power to detect interaction effects was governed primarily by the rate of uptake for the interaction effect, with contribution from the rate and structure of uptake for the main effects.

\begin{figure}
    \centering
    \includegraphics[width=1\linewidth]{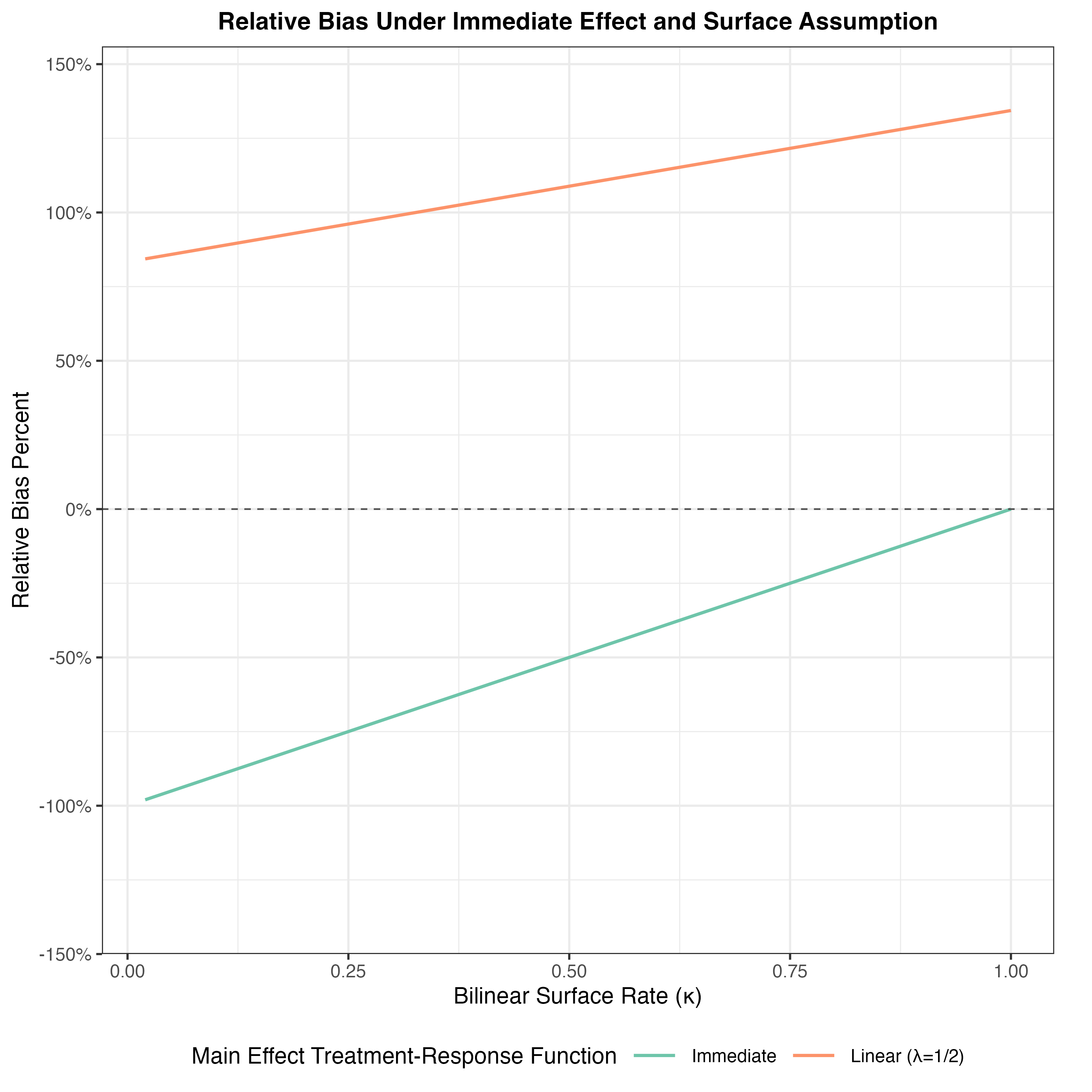}
    \caption[Relative Bias Under Immediate Effect Assumptions by Bilinear Surface Parameter]{Relative bias percent ($(\text{E}[\hat{\theta}]-\theta)/\theta*100$) by bilinear surface rate parameter ($\kappa$). Green lines represent power computed under an immediate effect assumption and orange linear ($\lambda=1/2$) treatment-response function for main effects. All results under a cross-sectional symmetric factorial multiple intervention SWD with compound symmetry ($\gamma=0.5$), $I=8$ cluster, $T=7$ periods, $n=50$ individuals per cluster-period, BPICC$=\rho_b=0.05$, WPICC$=\rho_w=0.2$, standardized main effect $d=0.7$, and standardized interaction effect $d=0.35$. Bonferroni corrections were not utilized for this analysis.}
    \label{fig:aim2_q5_kappa_bias}
\end{figure}

Figure \ref{fig:aim2_q5_kappa_bias} displays the relative bias of the interaction effect estimates when the data were generated under a bilinear interaction surface but analyzed under an immediate effect model. Under an immediate effect model for main effects, interaction estimates were strongly negatively biased for interaction effects with relative bias approaching $-100\%$ as $\kappa \rightarrow 0$ and increased monotonically as $\kappa \rightarrow 1$. In contrast, under linear main effects ($\lambda=1/2$), interaction effects were positively biased across all $\kappa$ with the magnitude of bias increasing as $\kappa$ increased. Thus we found that under bilinear interaction surfaces, the direction and magnitude of bias from assuming an immediate/constant effect surface for the interaction effect depends strongly on the main-effect uptake shape.

\subsection{Supplemental Materials C: Full Worked Example} 

Consider a hospital system implementing an artificial intelligence (AI)-based prescribing policy across its hospitals (clusters). \textit{Intervention 1} introduces an AI-driven risk flag that identifies patients at an increased risk for adverse drug interactions. As clinicians are likely to adopt and trust the tool gradually, we would expect a linear ramp-up in effectiveness as familiarity and confidence in the tool increase. \textit{Intervention 2} introduces an electronic health record (EHR) modification that sets an override-able constraint when a potentially conflicting prescription is ordered. This intervention would likely exhibit a latency period corresponding to configuration, training, and workflow integration, followed by a gradual increase in effectiveness. Finally, we would expect a positive interaction between the two interventions as clinicians will likely respond to the AI flag more effectively once the EHR safeguard is fully implemented and the systems reinforce the recommendation.

\begin{figure}
    \centering
    \includegraphics[width=0.8\linewidth]{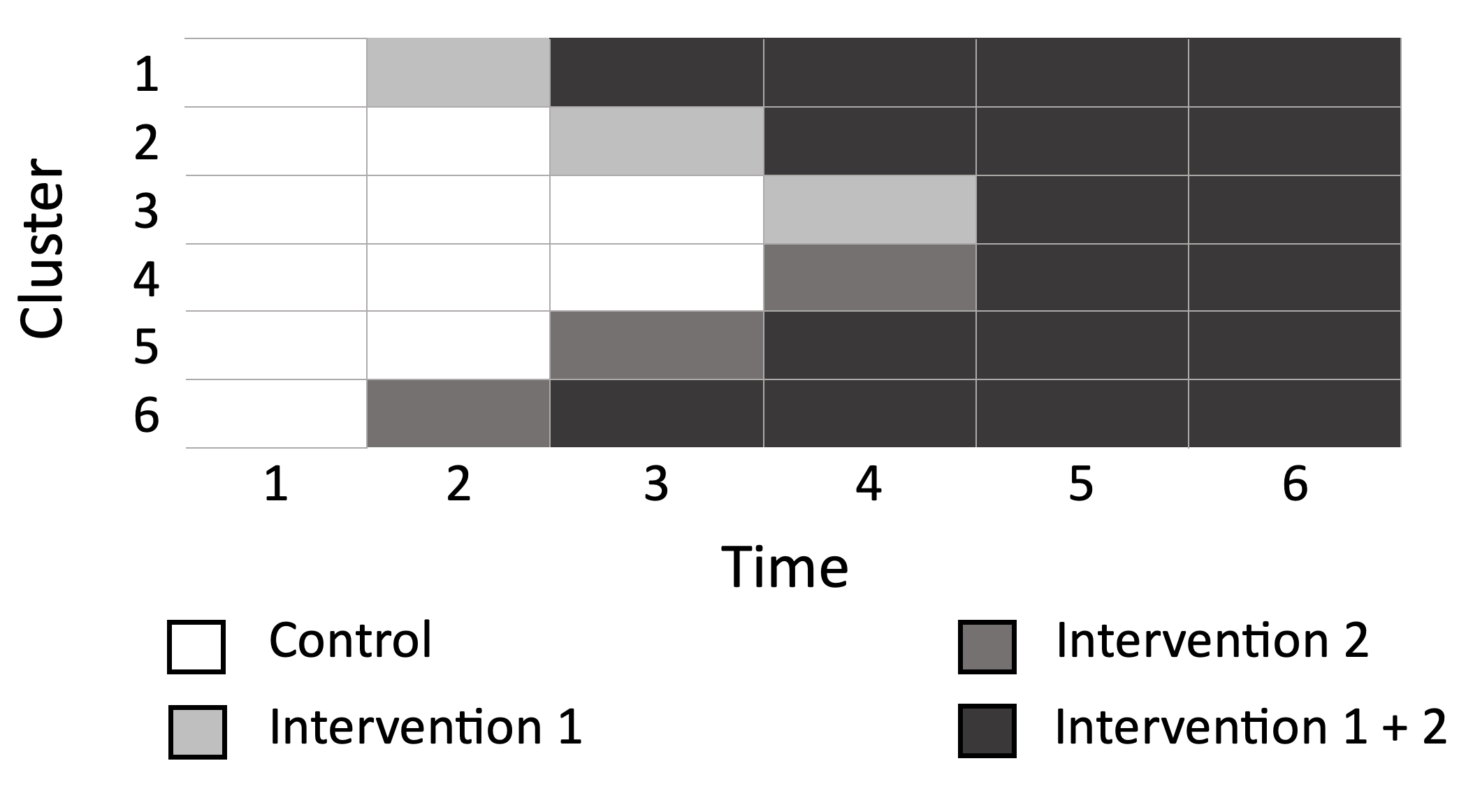}
    \caption[Example Study Design]{Example study design for 6 clusters across 6 periods.  White cluster period cells are in the control period, light gray are receiving only intervention 1, mid-gray are receiving only intervention 2, and dark gray receiving both interventions 1 and 2.}
    \label{tab:exstudydesign2}
\end{figure}

In this example, we consider a factorial design M-SWD with $I=6$ clusters, $J=6$ time periods, and 2 interventions with an assumed interaction and control arm. Under this example, intervention 1 will have a linear effect curve with slope $\lambda_1=1/3$ such that $f_1(s)=\lambda_1s_1=(1/3)s_1$, intervention 2 will have a piecewise linear effect curve with latency $L_2=1$, slope $\lambda_2=0.5$, and duration $\Delta_2=1/\lambda_2=2 $ such that 
\begin{equation*}
    f=\begin{cases} 0 & s_2 < L_2 \\
\lambda_2(s_2-L_2) & L_2 \leq s_2 \leq L_2+\Delta_2 \\
1 & s_2 > L_2 + \Delta_2 \end{cases} = 
\begin{cases} 0 & s_2 < 1 \\
0.5(s_2-1) & 1 \leq s_2 \leq 1+2=3 \\
1 & s_2 > 3 \end{cases} ,
\end{equation*} 
and the interaction term will have a bilinear surface with scalar $\kappa=0.9$ such that $f_{1,2}(s_1,s_2)= \kappa f_1(s_1)f_2(s_2)=0.9f_1(s_1)f_2(s_2)$.  The study design is shown in Figure \ref{tab:exstudydesign2}. 


For each vector, $s_q$ we can then calculate the full weight vector $h_q=\text{max}\{1,\int_{s_q=s_{qij}}^{s_{qij}+1} f_q(s_q) ds_q\}$ vectors:
\begin{align*}
    h_1(s_1) = \text{max}\{1,\lambda_1(s_{1}+1/2)\}&= \{\lambda_1(0+1/2), \lambda_1(1+1/2), \lambda_1(2+1/2),1, 1, 1\} \\
    &=\{\lambda_1(1/2), \lambda_1(3/2), \lambda_1(5/2), 1, 1,1\} \\
    &=\{0.167,0.5, 0.833, 1,1,1\}
\end{align*}
and
\begin{align*}
    h_2(s_2) &= \{0, \lambda_2(2-1), \lambda_2(3-1), \lambda_2(4-1), 1, 1\} \\
    &=\{0,0.25,0.75,1,1,1\},
\end{align*}
corresponding to exposure times $s \in \{0,5\}.$

Since $f_1(s_1)$ and $f_2(s_2)$ are independent, we can write $h_{1,2}$, as a $s_1 \times s_2$ matrix such that
\begin{align*}
h_{1,2}(s_1,s_2)&=\int_{s_2=s_{2ij}}^{s_{2ij}+1} \int_{s_1=s_{1ij}}^{s_{1ij}+1} f_{1,2}(s_1,s_2)ds_1ds_2 \\
&= \kappa \int_{s_1=s_{1ij}}^{s_{1ij}+1}\lambda_1s_1 ds_1 \int_{s_2=s_{2ij}}^{s_{2ij}+1}[\lambda_2(s_2-L_2)I(L_2 < s_2 \leq L_2+\Delta_2)+I(s_2>L_2+\Delta_2)]ds_2  \\
&= 0.9 \int_{s_1=s_{1ij}}^{s_{1ij}+1}(1/3)s_1 ds_1 \int_{s_2=s_{2ij}}^{s_{2ij}+1}[0.5(s_2-1)I(1 < s_2 \leq 3)+I(s_2>3)]ds_2  \\ \\
&= \begin{pmatrix}
\vspace{4pt}
   0    &  0.04  & 0.11   & 0.15   & 0.15 & 0.15 \\ \vspace{4pt}
   0  & 0.11 & 0.34 & 0.45 & 0.45 & 0.45 \\ \vspace{4pt}
  0 & 0.19 & 0.56 & 0.75 & 0.75 & 0.75 \\ \vspace{4pt}
   0  & 0.23 &0.68 & 0.9 &0.9 & 0.9\\ \vspace{4pt}
    0  & 0.23 & 0.68 & 0.9 & 0.9 & 0.9\\ \vspace{4pt}
   0  & 0.23 & 0.68 & 0.9 & 0.9 & 0.9   
\end{pmatrix}_{s_1 \times s_2} .
\end{align*}
We can then write the full design matrix $\mathbf{Z}$:

\[
\mathbf{Z}=
\left[
\begin{array}{ccccccccc}
        1 & 1 & 0 & 0 & 0 & 0 & 0&0 &0 \\
        1& 0 & 1 & 0 & 0 &0 &1 & 0  &0 \\
        1 & 0 & 0 & 1 & 0 & 0 & 1 & 1 & 1 \\
        1 & 0 & 0 & 0  & 1 &0 & 1& 1 & 1 \\
        1 & 0 & 0 & 0 & 0 &1 &1 & 1 & 1 \\
        1 & 0 &0 &0 &0 & 0 & 1 & 1 & 1 \\
        \hline
             1 & 1 & 0 & 0 & 0 & 0 & 0& 0& 0\\
        1& 0 & 1 & 0 & 0 &0 & 0 & 0 & 0 \\
        1 & 0 & 0 & 1 & 0 & 0 & 1 & 0 & 0 \\
        1 & 0 & 0 & 0  & 1 &0 & 1& 1& 1\\
        1 & 0 & 0 & 0 & 0 &1 & 1 & 1 &1 \\
        1 & 0 &0 &0 &0 & 0 & 1 & 1& 1\\
        \hline
             1 & 1 & 0 & 0 & 0 & 0 & 0 & 0 & 0 \\
        1& 0 & 1 & 0 & 0 &0 & 0 & 0& 0 \\
        1 & 0 & 0 & 1 & 0 & 0 & 0 &0 &0 \\
        1 & 0 & 0 & 0  & 1 &0 & 1& 0& 0\\
        1 & 0 & 0 & 0 & 0 &1 &1 & 1& 1\\
        1 & 0 &0 &0 &0 & 0 &1 &1 &1 \\
        \hline
             1 & 1 & 0 & 0 & 0 & 0 & 0 &0 &0 \\
        1& 0 & 1 & 0 & 0 &0 & 0 &1 &0 \\
        1 & 0 & 0 & 1 & 0 & 0 & 1 & 1 & 1 \\
        1 & 0 & 0 & 0  & 1 &0 & 1 &1  &1 \\
        1 & 0 & 0 & 0 & 0 &1 &1 &1 &1 \\
        1 & 0 &0 &0 &0 & 0 & 1& 1& 1\\
        \hline
             1 & 1 & 0 & 0 & 0 & 0 & 0 & 0 &0 \\
        1& 0 & 1 & 0 & 0 &0 & 0&0 &0 \\
        1 & 0 & 0 & 1 & 0 & 0 & 0& 1& 0 \\
        1 & 0 & 0 & 0  & 1 &0 & 1 & 1 & 1 \\
        1 & 0 & 0 & 0 & 0 &1 & 1 & 1 & 1\\
        1 & 0 &0 &0 &0 & 0 & 1& 1 &1 \\
        \hline
             1 & 1 & 0 & 0 & 0 & 0 & 0 & 0 & 0 \\
        1& 0 & 1 & 0 & 0 &0 & 0 & 0& 0\\
        1 & 0 & 0 & 1 & 0 & 0 &0 &0 &0 \\
        1 & 0 & 0 & 0  & 1 &0 & 0 & 1&  0\\
        1 & 0 & 0 & 0 & 0 &1 & 1 & 1 & 1 \\
        1 & 0 &0 &0 &0 & 0 & 1&1 & 1\\
\end{array}
\right]
\]
and the modified design matrix $\mathbf{Z}^*$:
\[
\mathbf{Z}^*=
\left[
\begin{array}{ccccccccc}
        1 & 1 & 0 & 0 & 0 & 0 & 0&0 &0 \\
        1& 0 & 1 & 0 & 0 &0 &0.167 & 0  &0 \\
        1 & 0 & 0 & 1 & 0 & 0 & 0.5 & 0 & 0 \\
        1 & 0 & 0 & 0  & 1 &0 & 0.833& 0.25 & 0.19 \\
        1 & 0 & 0 & 0 & 0 &1 &1 & 0.75 & 0.68 \\
        1 & 0 &0 &0 &0 & 0 & 1 & 1 & 0.9 \\
        \hline
             1 & 1 & 0 & 0 & 0 & 0 & 0& 0& 0\\
        1& 0 & 1 & 0 & 0 &0 & 0 & 0 & 0 \\
        1 & 0 & 0 & 1 & 0 & 0 & 0.167 & 0 & 0 \\
        1 & 0 & 0 & 0  & 1 &0 & 0.5& 0& 0\\
        1 & 0 & 0 & 0 & 0 &1 & 0.83 & 0.25 &0.19 \\
        1 & 0 &0 &0 &0 & 0 & 1 & 0.75& 0.68\\
        \hline
             1 & 1 & 0 & 0 & 0 & 0 & 0 & 0 & 0 \\
        1& 0 & 1 & 0 & 0 &0 & 0 & 0& 0 \\
        1 & 0 & 0 & 1 & 0 & 0 & 0 &0 &0 \\
        1 & 0 & 0 & 0  & 1 &0 & 0.167& 0& 0\\
        1 & 0 & 0 & 0 & 0 &1 &0.5 & 0& 0\\
        1 & 0 &0 &0 &0 & 0 &0.83 &0.25& 0.68 \\
        \hline
             1 & 1 & 0 & 0 & 0 & 0 & 0 &0 &0 \\
        1& 0 & 1 & 0 & 0 &0 & 0 &0 &0 \\
        1 & 0 & 0 & 1 & 0 & 0 & 0.167& 0.25 & 0.04 \\
        1 & 0 & 0 & 0  & 1 &0 & 0.5 &0.75  &  0.34 \\
        1 & 0 & 0 & 0 & 0 &1 &0.83 &1 &0.75 \\
        1 & 0 &0 &0 &0 & 0 & 1& 1& 0.9\\
        \hline
             1 & 1 & 0 & 0 & 0 & 0 & 0 & 0 &0 \\
        1& 0 & 1 & 0 & 0 &0 & 0&0 &0 \\
        1 & 0 & 0 & 1 & 0 & 0 & 0& 0& 0 \\
        1 & 0 & 0 & 0  & 1 &0 & 0.167 & 0.25 & 0.04 \\
        1 & 0 & 0 & 0 & 0 &1 & 0.5 & 0.75 & 0.34\\
        1 & 0 &0 &0 &0 & 0 & 0.833 & 1 & 0.75 \\
        \hline
             1 & 1 & 0 & 0 & 0 & 0 & 0 & 0 & 0 \\
        1& 0 & 1 & 0 & 0 &0 & 0 & 0& 0\\
        1 & 0 & 0 & 1 & 0 & 0 &0 &0 &0 \\
        1 & 0 & 0 & 0  & 1 &0 & 0 & 0&  0\\
        1 & 0 & 0 & 0 & 0 &1 & 0.167 & 0.25 & 0.04 \\
        1 & 0 &0 &0 &0 & 0 & 0.5&0.75 & 0.34\\
\end{array}
\right]
\]
where in both matrices column 7 corresponds to $X_1$, column 8 to $X_2$, and column 9 to $X_{1}X_2$.

We can define the difference of the fixed treatment effect matrices $\mathbf{D}=\mathbf{Z}^*-\mathbf{Z}$ as the discrepancy between the time-weight design and the assumed instantaneous design. When estimation is conducted using the full design matrix $\mathbf{Z}$ instead of the assumed matrix $\mathbf{Z}^*$, this misspecification impacts the GLS estimator including leading to biased treatment effects and variance inflation. Looking at the partitioned fixed treatment effect matrix $\mathbf{[D^{(1)} \hspace{6pt} \mathbf{D}^{(2)} \hspace{6pt} \mathbf{D}^{(1,2)}]} =[\mathbf{Z}^{*(1)} \hspace{6pt} \mathbf{Z}^{*(2)} \hspace{6pt} \mathbf{Z}^{*(1,2)}] - [\mathbf{Z}^{(1)} \hspace{6pt} \mathbf{Z}^{(2)} \hspace{6pt} \mathbf{Z}^{(1,2)}]$ for the above example, we find a notable design-weight deviation

\[
[\mathbf{D^{(1)}} \hspace{6pt} \mathbf{D}^{(2)} \hspace{6pt} \mathbf{D}^{(1,2)}] =
\left[
\begin{array}{ccc}
         0&0 &0 \\
        0.17 & 0  &0 \\
        0.5 & 0 & 0 \\
        0.83& 0.25 & 0.19 \\
        1 & 0.75 & 0.68 \\
         1 & 1 & 0.9 \\
        \hline
       0& 0& 0\\
       0 & 0 & 0 \\
        0.17 & 0 & 0 \\
        0.5& 0& 0\\
         0.83 & 0.25 &0.19 \\
         1 & 0.75& 0.68\\
        \hline
          0 & 0 & 0 \\
      0 & 0& 0 \\
       0 &0 &0 \\
      0.17& 0& 0\\
       0.5 & 0& 0\\
       0.83 &0.25&0.19 \\
        \hline
          0 &0 &0 \\
         0 &0 &0 \\
      0.17& 0.25 & 0.04 \\
        0.5 &0.75  &  0.34 \\
      0.83 &1 &0.75 \\
         1& 1& 0.9\\
        \hline
        0 & 0 &0 \\
        0&0 &0 \\
         0& 0& 0 \\
        0.17 & 0.25 & 0.04 \\
       0.5 & 0.75 & 0.34\\
        0.83 & 1 & 0.75 \\
        \hline
         0 & 0 & 0 \\
       0 & 0& 0\\
        0 &0 &0 \\
        0 & 0&  0\\
       0.17 & 0.25 & 0.04 \\
         0.5&0.75 & 0.34\\
    \end{array}
\right]
 - 
\mathbf{Z}=
\left[
\begin{array}{ccc}
        0&0 &0 \\
     1 & 0  &0 \\
         1 & 1 & 1 \\
        1& 1 & 1 \\
    1 & 1 & 1 \\
        1 & 1 & 1 \\
        \hline
          0& 0& 0\\
        0 & 0 & 0 \\
         1 & 0 & 0 \\
        1& 1& 1\\
       1 & 1 &1 \\
       1 & 1& 1\\
        \hline
        0 & 0 & 0 \\
        0 & 0& 0 \\
        0 &0 &0 \\
        1& 0& 0\\
        1 & 1& 1\\
        1 &1 &1 \\
        \hline
          0 &0 &0 \\
       0 &1 &0 \\
         1 & 1 & 1 \\
         1 &1  &1 \\
       1 &1 &1 \\
         1& 1& 1\\
        \hline
          0 & 0 &0 \\
        0&0 &0 \\
         0& 1& 0 \\
         1 & 1 & 1 \\
         1 & 1 & 1\\
         1& 1 &1 \\
        \hline
       0 & 0 & 0 \\
        0 & 0& 0\\
        0 &0 &0 \\
        0 & 1&  0\\
       1 & 1 & 1 \\
         1&1 & 1\\
    \end{array}
\right]
= 
\left[
\begin{array}{ccc}
         0&0 &0 \\
        -0.83 & 0  &0 \\
        -0.5 & -1 & -1 \\
        -0.17& -0.75 & -0.81 \\
        0 & -0.25 & -0.32 \\
         0 & 0 & -0.1 \\
        \hline
       0& 0& 0\\
       0 & 0 & 0 \\
        -0.83 & 0 & 0 \\
        -0.5& -1& -1\\
         -0.17 & -0.75 & -0.81 \\
         0 & -0.25& -0.32\\
        \hline
          0 & 0 & 0 \\
      0 & 0& 0 \\
       0 &0 &0 \\
      -0.83 & 0& 0\\
       -0.5 & -1& -1\\
       -0.17 &-0.75&-0.81 \\
        \hline
          0 &0 &0 \\
         0 &-1 &0 \\
      -0.83& -0.75 & -0.96 \\
        -0.5 & -0.25 &  -0.66 \\
      -0.17 &0 &-0.25 \\
         0& 0& -0.1\\
        \hline
        0 & 0 &0 \\
        0&0 &0 \\
         0& -1& 0 \\
        -0.83 & -0.75 & -0.96 \\
       -0.5 & -0.25 & -0.66\\
        -0.17 & 0 & -0.25 \\
        \hline
         0 & 0 & 0 \\
       0 & 0& 0\\
        0 &0 &0 \\
        0 & 0&  0\\
       -0.83 & -0.75 & -0.96 \\
         -0.5&-0.25 & -0.66\\
 \end{array}
\right]
\].

Importantly, we note that under the example structures, all observed values of the interaction experience some design-weight deviation. This example illustrates how distinct exposure-response parameterizations propagate into $\mathbf{Z}^*$, and consequently into bias and variance structures.

\subsection{Supplemental Materials D: Example Covariance Matrix Under Treatment-Effect Heterogeneity}

We assume a factorial design with $Q=2$ main effect interventions, plus an interaction term and control, such that we are looking to estimate $(\hat{\theta}_1, \hat{\theta}_2, \hat{\theta}_3)$. We include all previously specified random effects in our model, with a flexible $R$ cluster-period effect and random effect for treatment effect ($b_{qi}$). As we are focused on random effects, without loss of generality we assume a standard design matrix $\mathbf{Z}$ with no fixed treatment-time effect. The cluster-period mean model under these assumptions is:
\begin{equation}
    \bar{Y}_{ij} = \mu + \beta_j + \alpha_i +\nu_{ij} + \psi_{i.} +  [\theta_1+b_{1i}]X_{1ij} + [\theta_2+b_{2i}]X_{2ij}  + [\theta_3 +b_{3i}]X_{1ij}X_{2ij} + e_{ij.} .
\end{equation}

\begin{figure}
    \centering
    \includegraphics[width=0.6\linewidth]{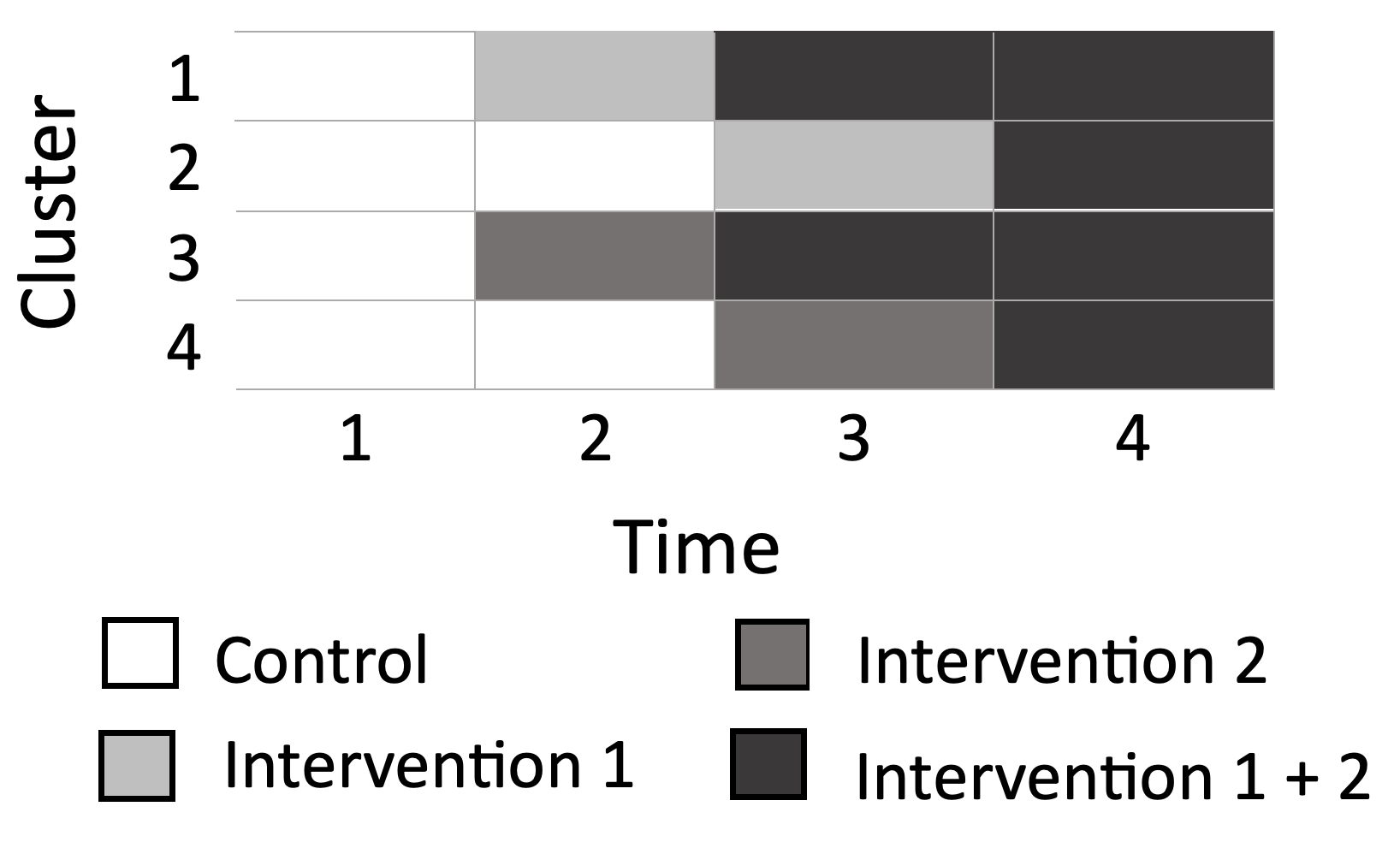}
    \caption[Example Study Design]{Example study design for 4 clusters across 4 periods.  White cluster period cells are in the control period, light gray are receiving only intervention 1, mid-gray are receiving only intervention 2, and dark gray receiving both interventions 1 and 2.}
    \label{tab:exstudydesign}
\end{figure}

The study design for this example is shown in Figure \ref{tab:exstudydesign}. Under this study design, we have fixed effect vector $\boldsymbol{\Theta} = (\mu, \beta_1, \beta_2, \beta_3, \theta_1, \theta_2,\theta_3)$. We can write the corresponding design matrix $\mathbf{Z}$:
\begin{equation}
    \mathbf{Z} = \begin{pmatrix}

        1 & 1 & 0 & 0 & 0 & 0 & 0 \\
        1 & 0 & 1 & 0 & 1 &0 & 0 \\
        1 & 0 & 0 & 1 & 1 & 1 & 1 \\
        1 & 0 & 0 & 0 & 1 & 1 & 1\\
    1 & 1 & 0 & 0 &0 & 0 & 0 \\
    1& 0 & 1 &0 & 0 & 0 & 0 \\
    1 & 0 & 0 & 1 & 1 & 0 & 0\\
    1 & 0 & 0 & 0 & 1 & 1 & 1 \\
        1 & 1 & 0 & 0 & 0 & 0 &0 \\
    1 & 0 & 1 & 0 &0 & 1 & 0 \\
    1 & 0 & 0 & 1& 1 & 1 & 1 \\
    1 & 0 & 0 & 0 & 1 & 1 & 1 \\
       1 & 1& 0 & 0 & 0 & 0 &0 \\
   1 & 0 & 1& 0  & 0 & 0 & 0\\
   1 & 0 & 0 & 1 & 0 & 1 & 0  \\
   1 & 0 & 0 & 0 & 1 & 1 & 1\\
    \end{pmatrix}
\end{equation}
where column 5 corresponds to $X_1$, column 6 to $X_2$, and column 7 to $X_{1}X_2$. We can write the covariance matrix, $V=\sigma^2_\nu R + \sigma^2_c I_T  + (\sigma^2_\alpha + \sigma^2_\zeta)J_T + \sum_{q=1}^Q\tau_qZ^{(q)}  = \sigma^2_\nu R + \sigma^2_c I_T + (\sigma^2_\alpha + \sigma^2_\zeta)J_T + \tau_1Z^{(1)} + \tau_2Z^{(2)} + \tau_3Z^{(1,2)}  $, as
\begin{equation*}
    \mathbf{V} = \begin{pmatrix}
        V_1 & 0 & 0 & 0 \\
        0 & V_2 & 0 & 0 \\
        0 & 0 & V_3 & 0 \\
        0 & 0 & 0 & V_4
    \end{pmatrix} .
\end{equation*}
Let $k = \sigma^2_\alpha + \sigma^2_\zeta$. Then 
\begin{equation*}
    V_1 = \begin{pmatrix}
        k + \sigma^2_\nu + \sigma^2_c & k + \sigma^2_\nu r_{12} & k + \sigma^2_\nu r_{13} & k + \sigma^2_\nu r_{14} \\
        
        k + \sigma^2_\nu r_{21} & k + \sigma^2_\nu + \sigma^2_c + \tau_1^2 & k + \sigma^2_\nu r_{23} + \tau_1^2  & k + \sigma^2_\nu r_{24} + \tau_1^2  \\
        
        k + \sigma^2_\nu r_{31} & k + \sigma^2_\nu r_{32} + \tau_1^2  & k + \sigma^2_\nu + \sigma^2_c  + \tau_1^2 + \tau^2_2 + \tau_3^2  & k + \sigma^2_\nu r_{34} + \tau_1^2 + \tau_2^2 + \tau^2_3\\
        
        k + \sigma^2_\nu r_{41} & k + \sigma^2_\nu r_{42} + \tau_1^2  & k + \sigma^2_\nu r_{43} + \tau_1^2 + \tau^2_2 + \tau^2_3& k + \sigma^2_\nu + \sigma^2_c  + \tau_1^2 + \tau^2_2  + \tau^2_3
    \end{pmatrix}
\end{equation*}

\begin{equation*}
    V_2 = \begin{pmatrix}
        k + \sigma^2_\nu + \sigma^2_c & k + \sigma^2_\nu r_{12} & k + \sigma^2_\nu r_{13} & k + \sigma^2_\nu r_{14} \\
        
        k + \sigma^2_\nu r_{21} & k + \sigma^2_\nu + \sigma^2_c & k + \sigma^2_\nu r_{23}   & k + \sigma^2_\nu r_{24}  \\
        
        k + \sigma^2_\nu r_{31} & k + \sigma^2_\nu r_{32}   & k + \sigma^2_\nu + \sigma^2_c  + \tau_1^2  & k + \sigma^2_\nu r_{34} + \tau_1^2 \\
        
        k + \sigma^2_\nu r_{41} & k + \sigma^2_\nu r_{42}  & k + \sigma^2_\nu r_{43} + \tau_1^2  & k + \sigma^2_\nu + \sigma^2_c  + \tau_1^2 + \tau_2^2 + \tau^2_3
    \end{pmatrix}
\end{equation*}

\begin{equation*}
    V_3 = \begin{pmatrix}
        k + \sigma^2_\nu + \sigma^2_c & k + \sigma^2_\nu r_{12} & k + \sigma^2_\nu r_{13} & k + \sigma^2_\nu r_{14} \\
        
        k + \sigma^2_\nu r_{21} & k + \sigma^2_\nu + \sigma^2_c + \tau_2^2 & k + \sigma^2_\nu r_{23} + \tau_2^2  & k + \sigma^2_\nu r_{24} + \tau_2^2  \\
        
        k + \sigma^2_\nu r_{31} & k + \sigma^2_\nu r_{32} + \tau_2^2  & k + \sigma^2_\nu + \sigma^2_c  + \tau_2^2  + \tau_1^2 + \tau^2_3 & k + \sigma^2_\nu r_{34} + \tau_2^2 + \tau_1^2 + \tau_3^2\\
        
        k + \sigma^2_\nu r_{41} & k + \sigma^2_\nu r_{42} + \tau_2^2   & k + \sigma^2_\nu r_{43} + \tau_2^2 + \tau^2_1 + \tau^2_3  & k + \sigma^2_\nu + \sigma^2_c  + \tau_2^2 + \tau^2_1 + \tau^2_3
    \end{pmatrix}
\end{equation*}

\begin{equation*}
    V_4 =\begin{pmatrix}
        k + \sigma^2_\nu + \sigma^2_c & k + \sigma^2_\nu r_{12} & k + \sigma^2_\nu r_{13} & k + \sigma^2_\nu r_{14} \\
        
        k + \sigma^2_\nu r_{21} & k + \sigma^2_\nu + \sigma^2_c  & k + \sigma^2_\nu r_{23}   & k + \sigma^2_\nu r_{24}  \\
        
        k + \sigma^2_\nu r_{31} & k + \sigma^2_\nu r_{32}  & k + \sigma^2_\nu + \sigma^2_c  + \tau_2^2  & k + \sigma^2_\nu r_{34} + \tau_2^2 \\
        
        k + \sigma^2_\nu r_{41} & k + \sigma^2_\nu r_{42}   & k + \sigma^2_\nu r_{43} + \tau_2^2  & k + \sigma^2_\nu + \sigma^2_c  + \tau_2^2  + \tau^2_1 + \tau^2_3
    \end{pmatrix}
\end{equation*}
We note that under this structure, the $V_i$s are not uniform and thus the full $\mathbf{V}$ matrix is not block diagonal. Therefore, under treatment level heterogeneity, $\mathbf{V}$ cannot be inverted using standard principles, except under the case where $\tau_1^2=\tau^2_2=\tau^2_3=0$, resulting in no treatment level random effects. To compute Var($\hat{\theta}$) terms for power calculations under this structure, it is necessary to numerically compute $\text{Cov}(\hat{\boldsymbol{\Theta}})=(\mathbf{Z}^T\mathbf{V}^{-1}\mathbf{Z})^{-1}$ where $\text{Var}(\hat{\theta}_1)$ is the $[T+1, T+1]$ element of $\text{Cov}(\hat{\boldsymbol{\Theta}})$, $\text{Var}(\hat{\theta}_2)$ is the $[T+2,T+2]$ element of $\text{Cov}(\hat{\boldsymbol{\Theta}})$, and $\text{Var}(\hat{\theta}_3)$ is the $[T+3, T+3]$ element of $\text{Cov}(\hat{\boldsymbol{\Theta}})$.

\subsection{Supplemental Materials E: Validation of Theoretical Results}\label{sec:aim2val_details}

\begin{figure}
    \centering
    \includegraphics[width=1\linewidth]{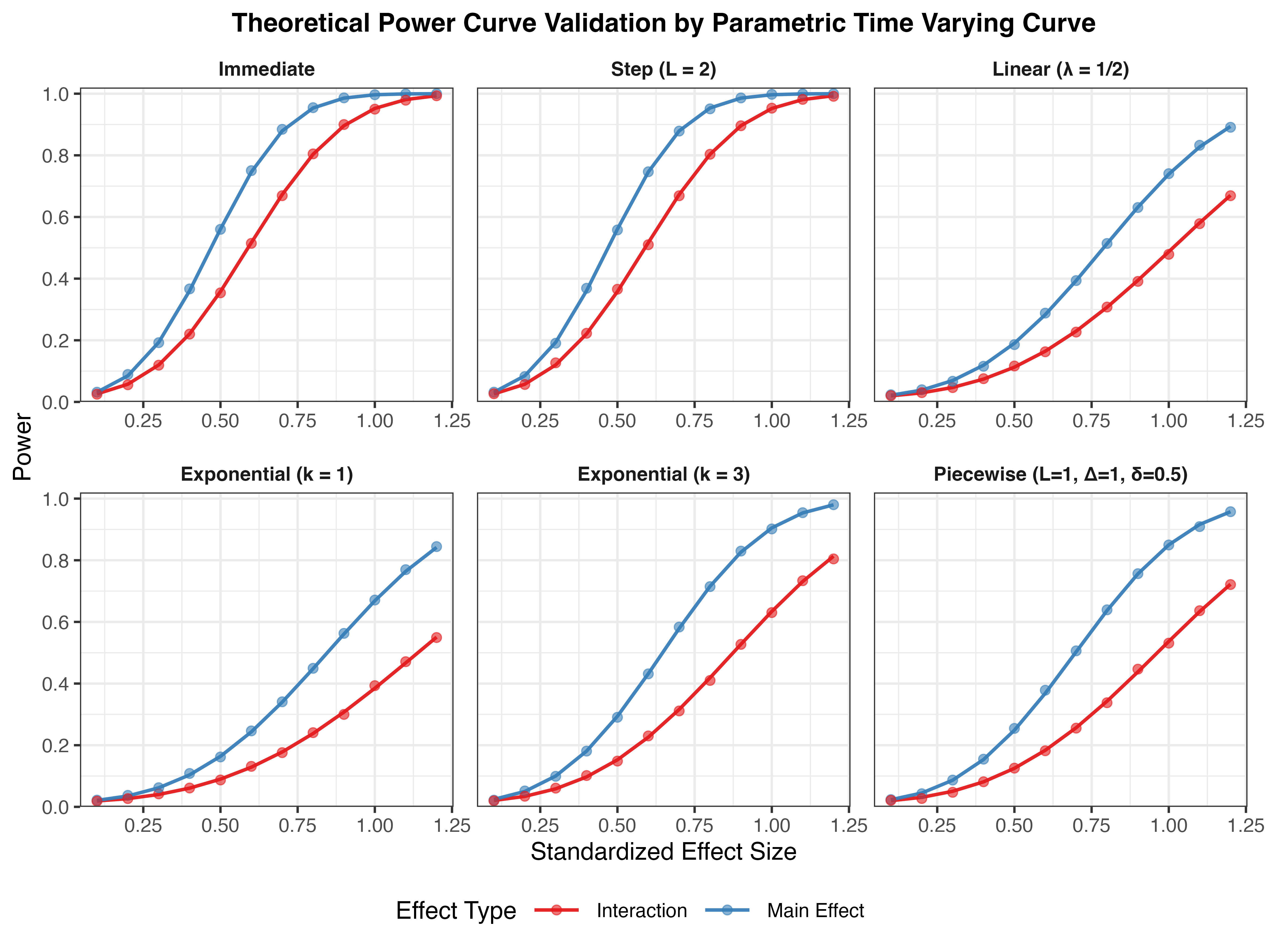}
    \caption[Aim 2 Validation]{Theoretical (lines) and simulated GLS-oracle (points) power curves by parametric time varying treatment-response functions. Results shown under a factorial design with $I=8$ clusters, $T=7$ periods, $n=50$ individuals per cluster period, $\rho_b=0.05$ between period ICC, and $\rho_w=0.2$ within period ICC. Compound Symmetry cluster-period correlation under moderate correlation ($\gamma=0.5$) was utilized. Blue points and lines correspond to main effects, with intervention 1 and intervention 2 collapsed due to symmetry, and red points and lines correspond to interaction effects. }
    \label{fig:aim2q01}
\end{figure}

\begin{figure}
    \centering
    \includegraphics[width=1\linewidth]{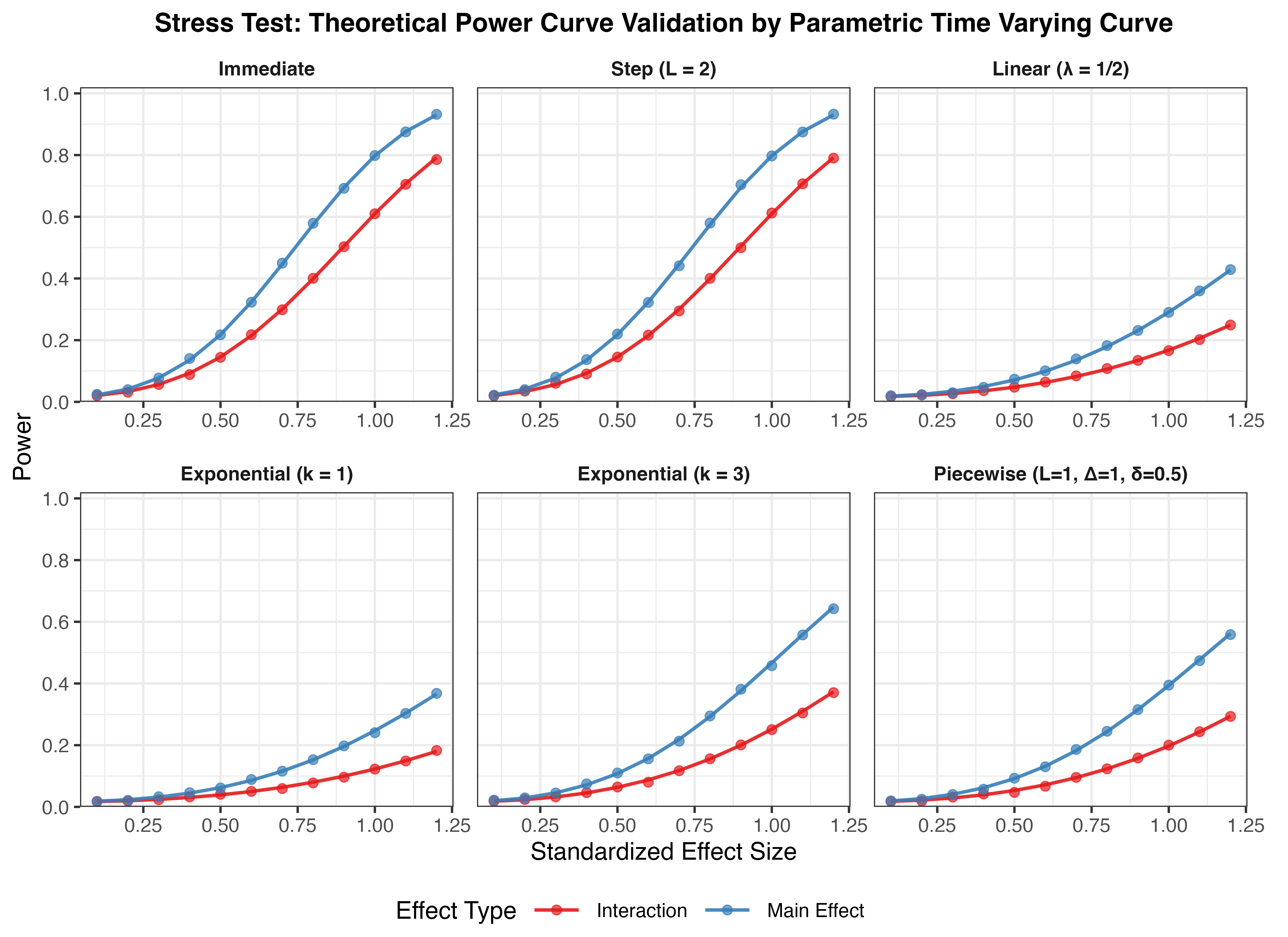}
    \caption[Aim 2 Validation]{Theoretical and simulated GLS-oracle power curves by parametric time varying treatment-response functions. Results shown under a factorial design with $I=6$ clusters, $T=6$ periods, $n=15$ individuals per cluster period, $\rho_b=0.2$ between period ICC, and $\rho_w=0.4$ within period ICC. Compound Symmetry cluster-period correlation under moderate correlation ($\gamma=0.5$) was utilized. Points indicate GLS-oracle simulations and lines represent theoretical power curves. Blue points and lines correspond to main effects, with intervention 1 and intervention 2 collapsed due to symmetry, and red points and lines correspond to interaction effects. }
    \label{fig:aim2q0_stress}
\end{figure}

In this section, we utilized simulations to validate the partial fixed effect design matrix theory and its resulting Wald-based power calculations under finite-sample M-SWDs. Figure \ref{fig:aim2q01} demonstrates near-perfect alignment between theoretical and simulated power curves for a representative design scenario, indicating that the asymptotic approximations were accurate in sample sizes typical of pragmatic M-SWDs.

Across all effect sizes and correlation structures within this scenario, mean differences between simulated and theoretical power were small (average difference [$\text{mean}(\text{Power}_\text{Simulated} -\text{Power}_{\text{Theoretical}})$] $<0.001$; average absolute difference [$\text{mean}(|\text{Power}_\text{Simulated} -\text{Power}_{\text{Theoretical}}|)$] $0.003$), with a maximum absolute difference of $0.012$. Percent differences [$\frac{\text{Power}_\text{Simulated}-\text{Power}_{\text{Theoretical}}}{\text{Power}_{\text{Theoretical}}}*100$] center near zero (mean = $<0.001$) with a range from $-12.1$ to $9.3$. As expected, larger percent deviations occurred primarily where power was near zero, while absolute differences remained negligible across the full power range. We also tested the alignment under more extreme conditions, and validation results were comparable to those described above. Further details can be found in Figure \ref{fig:aim2q0_stress}.

\subsection{Supplemental Materials F: Power Under Incomplete Designs with Additional Clusters}
\begin{figure}
    \centering
    \includegraphics[width=0.95\linewidth]{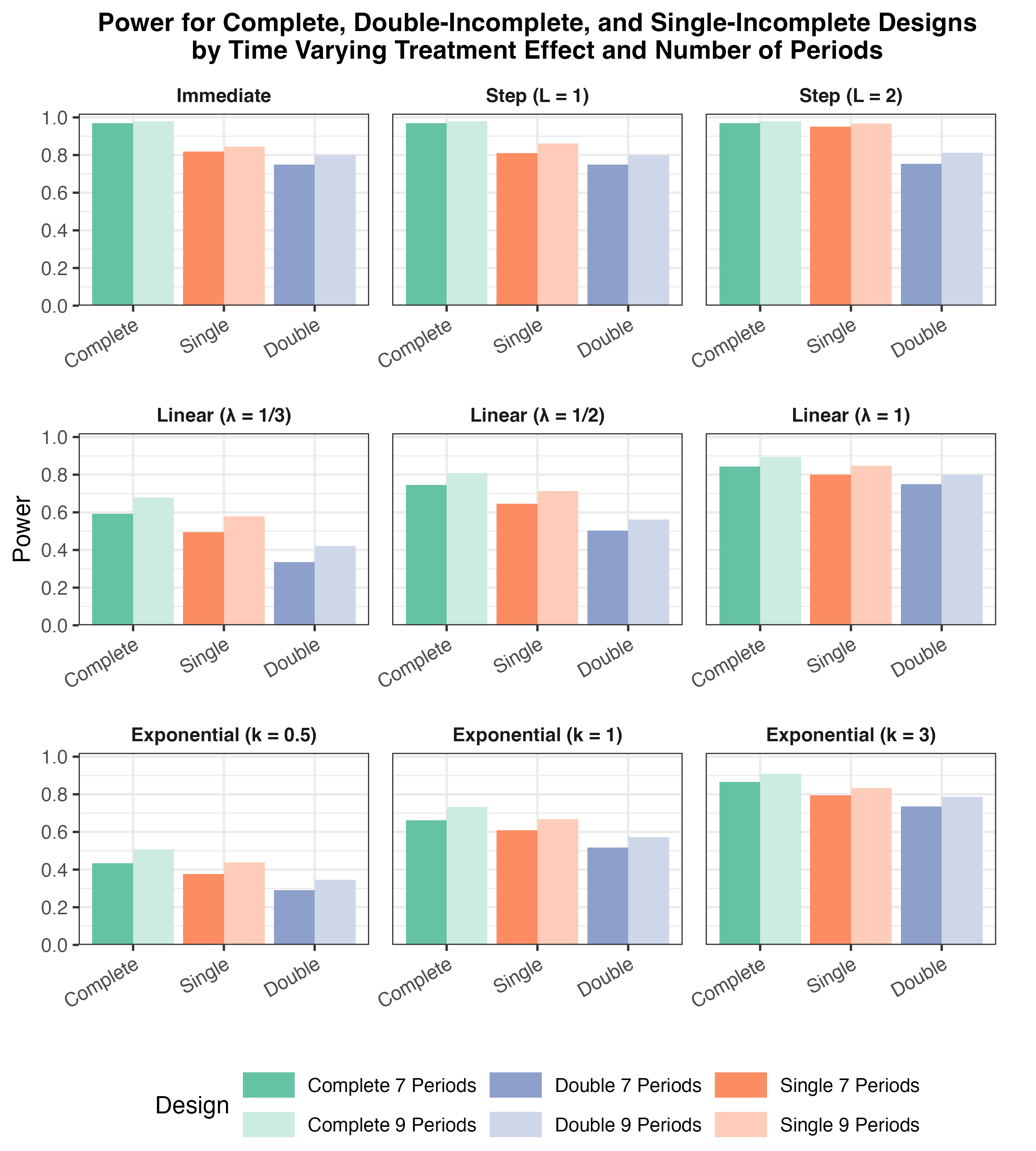}
    \caption[Power for Incomplete Designs by Time Varying Treatment Effect and Number of Periods]{Power by incomplete design type and time varying treatment effect across 7 and 9 periods. Bars show power under (i) complete design (green), (ii) single incomplete design that excludes the post-transition cluster-period for the first intervention received by the cluster (orange), and (iii) double-incomplete design that excludes the post-transition period for each intervention transition (blue). Darker colors indicate 7 periods as shown throughout this work and lighter colors represent 9 periods. All results under a cross-sectional symmetric factorial multiple intervention SWD with $I=8$ cluster, $T=7$ periods, $n=50$ individuals per cluster-period, BPICC$=\rho_b=0.05$, WPICC$=\rho_w=0.2$ and standardized maximum realized main effect sizes of $d=0.7$ and maximum standardized interaction effect of $d=0.35$. Power is computed under a correctly specified time-varying treatment-response model.}
    \label{fig:aim2_q4_appendix}
\end{figure}

Since power is reduced under incomplete designs, Figure \ref{fig:aim2_q4_appendix} demonstrates the impact of adding two additional periods on power under time-varying treatment effects for complete and incomplete designs. While both single and double incomplete designs were able to make modest power gains under the addition of two more periods, power to detect main effects remained lower for single and double incomplete designs under 9 periods than for complete designs with 7 periods. This finding demonstrates that the cost of information lost from the pivotal first post-transition periods is too significant to be able to be offset from the addition of two periods at the end of the study. 

\end{document}